\documentclass[trackchanges, twocolumn]{aastex701}

\usepackage{amsmath}
\usepackage[normalem]{ulem}
\usepackage{xcolor}

\begin{document}

\title{Tracking Large-Amplitude Oscillations in a Partially Erupting Quiescent Prominence: Evolution of Oscillation Properties Before and After Eruption}

\correspondingauthor{S. Krishna Prasad}
\email{krishna.prasad@aries.res.in}

\author[orcid=0009-0000-8577-8337,gname='Mansi']{Mansi}

\affiliation{Aryabhatta Research Institute of Observational Sciences(ARIES), Manora Peak, Nainital-263001, Uttarakhand, India}
\affiliation{Department of Physics, Indian Institute of Technology Roorkee, Roorkee-247667, Uttarakhand, India}
\email{venimansi15@gmail.com}  

\author[orcid=0000-0002-0735-4501,gname='S. Krishna', sname='Prasad']{S. Krishna Prasad} 

\affiliation{Aryabhatta Research Institute of Observational Sciences(ARIES), Manora Peak, Nainital-263001, Uttarakhand, India}
\email{krishna.prasad@aries.res.in}

\author[0000-0002-0687-6172]{Dmitrii Y. Kolotkov}
\affiliation{Centre for Fusion, Space and Astrophysics, Department of Physics, University of Warwick, Coventry CV4 7AL, UK}
\affiliation{Engineering Research Institute \lq\lq Ventspils International Radio Astronomy Centre (VIRAC) \rq\rq, Ventspils University of Applied Sciences, Ventspils, LV-3601, Latvia}
\email{D.Kolotkov.1@warwick.ac.uk}

\author[0000-0002-3841-313X]{Manuel Luna}
\affiliation{Departament de Física, Universitat de les Illes Balears (UIB), E-07122 Palma de Mallorca, Spain}
\affiliation{Institute of Applied Computing \& Community Code (IAC3), UIB, E-07122 Palma de Mallorca, Spain}
\email{manuel.luna@uib.es}

\author[0000-0001-8232-6626,sname=Pandey,gname=Ambrish]{Ambrish Pandey  }

\affiliation{Department of Physics, Indian Institute of Technology Roorkee, Roorkee-247667, Uttarakhand, India}
\email{ambrish.pandey@ph.iitr.ac.in}


\begin{abstract}
Solar prominences are dense, cool plasma structures suspended in the corona by magnetic fields, and can exhibit oscillatory motions as well as evolve into eruptive events associated with coronal mass ejections. In some cases, large-amplitude oscillations have been detected during the pre-eruptive and eruptive phases of a prominence, suggesting a possible link between oscillatory behavior and the destabilization of its supporting magnetic structure. However, the continuous evolution of oscillation properties across this transition, tracked in the same prominence material, remained poorly constrained. Using the SUVI onboard GOES, together with the FSI onboard Solar Orbiter, we analyzed a prominence observed on 2024 February 9 that exhibited large-amplitude transverse horizontal oscillations, underwent a slow rise, and subsequently bifurcated into two components, with the upper segment erupting as a coronal mass ejection confirmed in LASCO C2 onboard SOHO. A co-moving slit was used to track the same plasma parcel continuously in height, enabling a direct, single-parcel comparison of the oscillation properties before and after eruption. We found that the oscillation period increased from $\sim34$ to $\sim54~\mathrm{min}$, the initial displacement amplitude decreased from $\sim9$ to $\sim7~\mathrm{Mm}$, the damping time increased from $\sim77$ to $\sim134~\mathrm{min}$, and the characteristic velocity amplitude decreased from $\sim$29 to $\sim$13~km~s$^{-1}$. We further employed an analytic model to interpret this trend, finding it consistent with a weakening restoring force.
\end{abstract}



\section{Introduction} 
\label{intro}

 Solar prominences (when observed on the limb) or filaments (when observed on the disk) are giant plasma structures suspended in the solar corona, with temperatures of $T \approx 10^4$~K and electron densities of $10^9$-$10^{11}$~cm$^{-3}$ \citep{Labrosse_2010}. They are located above polarity inversion lines (PILs), where the radial component of the photospheric magnetic field changes sign \citep{Mackay2010}. Because prominence plasma is cooler and denser than the surrounding corona, it must be supported against gravity. This support is provided by magnetic dips, i.e., local spatial minima of the field lines where the upward-directed magnetic tension balances the downward gravitational pull \citep{Kippenhahn1957}. These concave-upward configurations typically exist within large-scale sheared arcades or magnetic flux ropes \citep[e.g.,][]{Mackay2010, Gibson2018}. \\Prominences are commonly divided, based on their location and magnetic environment, into two broad classes, quiescent and active region prominences. Quiescent prominences are large, long-lived structures located in quiet-Sun regions, with  lifetimes of days to several months \citep{Labrosse_2010}, and average magnetic field strengths of roughly 8-10~G \citep{Parenti2014}. Conversely, active region prominences are typically shorter-lived, lasting from minutes to hours \citep{Parenti2014}, and are embedded in stronger, more complex magnetic fields, with strengths of a few hundred gauss \citep{kuckein2009}. \\ Prominences are also highly dynamic structures that exhibit a wide range of oscillatory motions. The first systematic study of prominence oscillations was performed by \citet{Ramsey1965}, who observed periodic disappearance and reappearance of solar filaments in H$\alpha$ line-center images, later termed winking filaments.

Based on the velocity amplitudes of these oscillations, they can be categorized as large and small amplitude oscillations. Large-amplitude oscillations are typically associated with energetic disturbances and involve the motion of a substantial fraction, or even the entirety of the prominence, thereby representing a global response, with characteristic velocity amplitudes exceeding $10\ \mathrm{km\ s^{-1}}$. In contrast, small-amplitude oscillations are generally confined to localized regions of the prominence,  and exhibit lower velocity amplitudes, typically below $\sim 10\ \mathrm{km\ s^{-1}}$\citep{Arregui2018}.

Large-amplitude oscillations can manifest either as longitudinal motions of the prominence material along its axis (e.g., \citealt{Luna2014,Jing2003}) or as transverse displacements of the prominence axis, occurring in horizontal (e.g., \citealt{Hershaw2011}) or vertical directions (e.g., \citealt{Ramsey1965}) with respect to the solar surface. These oscillations are of particular importance as they involve the bulk motion of the structure and can be described using physical models  (e.g.,\citealt{Hyder1966,Kleczek1969,Luna2012}). By analyzing observable quantities such as the oscillation period, amplitude, and damping time, key physical parameters including the magnetic field strength, radius of curvature of the dip (where prominence plasma is present) can be inferred (e.g., \citealt{Shen2014,Vrsnak2007, Luna2014,Baweja2026}). In this way, large-amplitude oscillations provide a powerful means of performing prominence seismology, offering insights into plasma and magnetic conditions that are otherwise difficult to measure directly at such large coronal heights. 

In addition, large-amplitude oscillations are often observed in association with highly dynamic phases of prominence evolution, and in some cases, they are detected prior to or during eruptive events, suggesting a possible link between oscillatory behavior and the destabilization of the prominence structure. Observations have shown that prominences can undergo sustained oscillatory motions during their slow-rise or pre-eruptive phase, with oscillations persisting for several hours before eruption (e.g., \citealt{Isobe2006,Chen2008,Joshi2016,Foullon2009}). 

Oscillatory behavior during eruption has also been reported using spectroscopic observations e.g., \cite{Bocchialini2011}, who analyzed time series of spectral lines (He~I and Mg~X) and reported clear intensity and velocity oscillations during eruptive phases. 

While previous studies have investigated oscillatory properties either during pre-eruptive or eruptive phases, to our knowledge, the continuous tracking of the same plasma oscillations throughout its full evolution from a stable state, through the slow-rise phase, and into eruption has not yet been reported. A potential benefit of such tracking is to provide additional 
constraints to existing prominence models, especially by ensuring their validity through a comparison of oscillation behavior during the eruption phase. In this work, we employ a co-moving horizontal slit to follow a discrete plasma parcel, allowing us to characterize transverse horizontal oscillations and determine their properties both before and after the onset of eruption. To further contextualize our findings for the pre and post-eruption phases, we utilize the model developed by \cite{Dmitrii2018}. Through this framework, we qualitatively relate the observed prominence's transverse oscillation characteristics directly to the geometry and stability of its supporting magnetic dip, providing a physical basis to interpret the observed changes in the oscillation period.

The structure of this paper is as follows. In Section~\ref{data}, we describe the observations and data used in this study. Section~\ref{sec:analysis} outlines the methodology employed to track the prominence evolution and extract oscillation parameters, including the theoretical model used to interpret their evolution. The results are presented in Section~\ref{sec:results}, followed by a discussion interpreting the findings in terms of this model in Section~\ref{sec:discussion}. Finally, we summarize our conclusions in Section~\ref{sec:conclusions}.

\section{Data and Observations}
\label{data}

This study analyzes a prominence oscillation event on 2024 February 9, in which large-amplitude transverse oscillations were triggered following the impact of an EUV wave front at approximately 13:09~UT. The event was primarily observed using the Solar Ultraviolet Imager \cite[SUVI;][]{Darnel2022} onboard the Geostationary Operational Environmental Satellite (GOES-16). This prominence (shown in Figure \ref{fig:slit}) was previously analyzed by \citet{Zhang2024}, who reported large-amplitude transverse oscillations using the Atmospheric Imaging Assembly \citep[AIA;][]{Lemen2012} onboard the Solar Dynamics Observatory (SDO). While AIA provides high-resolution imaging of the low corona, its field of view limits the ability to follow the prominence evolution to larger heights, and consequently their study focused primarily on the early, stable phase. Here, we make use of the extended field of view and continuous temporal coverage of SUVI to track the same prominence material continuously from this early phase through the slow-rise, bifurcation into the subsequent eruptive rise, allowing us to report the evolution of its oscillatory properties across the full pre-eruptive to eruptive transition.

SUVI provides full-disk extreme ultraviolet (EUV) observations of the solar atmosphere in six wavelength channels (94, 131, 171, 195, 284, and 304~\AA), covering plasma temperatures from the chromosphere to the corona. A key advantage of SUVI is its extended field of view, reaching up to $\sim$1.6~$R_\odot$, which enables continuous tracking of prominences to greater heights. The data used in this study are Level~2 304~\AA images \citep{Darnel2018}, with a pixel scale of 2.5\arcsec\ ($\approx$ 1.8 Mm at the GOES-16 distance of $\approx$ 0.98 AU) and a cadence of 4~minutes. The temperature response of the 304~\AA\ channel exhibits multiple peaks demonstrating sensitivity at different temperatures ranging from $\log T \approx 4.9 \-- 6.2$, though it is dominated by \ion{He}{2} emission with its primary peak at $\log T \approx 4.9$ \citep[see Figure~1 and Table~4 of][]{Darnel2022}.

To further extend the observational coverage to larger coronal heights, we additionally utilize EUV observations from the Full Sun Imager (FSI) 304~\AA\ channel of the Extreme Ultraviolet Imager \citep{Rochus2020} onboard the Solar Orbiter \citep{Muller2020}. FSI provides an extended field of view reaching up to $\sim6~R_{\odot}$, for this event on 2024 February 9. This extended coverage is particularly important for the present event, as it allows us to follow the large-scale evolution of the prominence into the outer corona. The FSI 304 \AA\ images used in this study are Level 2 (L2) data from the \href{https://www.sidc.be/EUI/data/releases/202301_release_6.0/}{EUI Data Release 6.0} \citep{EUI_DR6}, with a temporal cadence of 6 minutes and a pixel scale of $4.44\arcsec$ ($\approx$ 2.7 Mm). On 2024 February 9, Solar Orbiter was located at a heliocentric distance of $\sim0.84$~AU, with a longitudinal separation of $\sim26.6^\circ$ from Earth, providing a complementary vantage point for observing the erupting prominence. While the transverse oscillatory signatures are more clearly resolved in the higher-resolution SUVI observations, which form the primary dataset for this study, FSI provides essential large-scale context, enabling the tracking of the prominence evolution to greater heights beyond the SUVI field of view.

We additionally use 171, 193, and 211~\AA\ channels of AIA to carefully examine the interaction between the prominence and the two EUV wave fronts reported in the earlier study of this event by \citet{Zhang2024}. Of these three channels, 193~\AA\ (Fe~\textsc{xii}, $\sim$1.5~MK) and 211~\AA\ (Fe~\textsc{xiv}, $\sim$2~MK) provide the clearest contrast required for this purpose, whereas 171~\AA\ (Fe~\textsc{ix}, $\sim$0.6~MK) is comparatively noisier. We therefore use the AIA 193~\AA\ channel for the analysis in Section~\ref{sec:forced}. We also use the AIA 304~\AA\ channel, at a pixel scale of 0.6$^{\prime\prime}$ and a cadence of 12~s, a substantial improvement over the 4~minute cadence of SUVI, to more precisely determine the time at which the prominence visibly separates into two segments.

We particularly analyze SUVI 304~\AA\ observations spanning from 12:41~UT to 19:30~UT for tracking the evolution of the prominence. The evolving prominence structure throughout the observed event is shown via sample images and accompanying animation in Figure~\ref{fig:moving_slit}. As can be seen, during its evolution, the prominence undergoes a clear structural separation into two components. Following this bifurcation, the upper segment continues to rise, while the lower portion partially drains back toward the solar surface. In this study, we focus on the upper segment, which continues the transverse oscillatory motion and subsequently forms the erupting component.

To confirm the presence of an associated coronal mass ejection (CME), we examined white-light coronagraph observations from the Large Angle and Spectrometric Coronagraph (LASCO; \citealt{Brueckner1995}) C2 instrument onboard the Solar and Heliospheric Observatory (SOHO), where a corresponding CME is clearly detected.

We identify the bifurcation point as marking the onset of eruption for the upper segment, since at this point it detaches from the underlying photospheric footpoints and begins its unbound rise. This is further supported by the extended FSI tracking of the same segment and its direct spatial continuity with the associated CME in LASCO C2 (detailed in Section~\ref{subsec:fsi_check}), which show the upper segment ascending as a single coherent, disconnected structure rather than one still tied to the solar surface. Table~\ref{tab:timeline} summarizes the phases of the event, their timings, and the instruments used.

\begin{deluxetable*}{lcll}
\tabletypesize{\small}
\tablecaption{Timeline and Evolutionary Phases of the 2024 February 9 Prominence Oscillation event\label{tab:timeline}}
\tablewidth{0pt}
\tablehead{
\colhead{Evolutionary Phase / Event} &
\colhead{Time (UT)} &
\colhead{Instrument(s)} &
\colhead{Observational Signatures and Dynamics}
}
\startdata
\parbox[t]{3.8cm}{\raggedright Quasi-static equilibrium} & 12:41--13:09 & \parbox[t]{3.4cm}{\raggedright SUVI 304~\AA} & \parbox[t]{7cm}{\raggedright Prominence at nearly constant height.} \\
\parbox[t]{3.8cm}{\raggedright EUV wave interaction} & 13:09--$\approx$13:30 & \parbox[t]{3.4cm}{\raggedright AIA 193~\AA, SUVI 304~\AA} & \parbox[t]{7cm}{\raggedright Successive arrivals of WF1 (13:09~UT) and WF2 (13:17~UT) drive a compound forced response of the prominence.} \\
\parbox[t]{3.8cm}{\raggedright Pre-eruption oscillation} & $\approx$13:30--14:17 & \parbox[t]{3.4cm}{\raggedright SUVI 304~\AA} & \parbox[t]{7cm}{\raggedright Transverse oscillation, free of direct EUV-wave forcing, lasting until the bifurcation.} \\
\parbox[t]{3.8cm}{\raggedright Onset of vertical rise} & $\approx$13:33 & \parbox[t]{3.4cm}{\raggedright SUVI 304~\AA} & \parbox[t]{7cm}{\raggedright Prominence height begins to increase.} \\
\parbox[t]{3.8cm}{\raggedright Bifurcation (eruption onset)} & 14:17 & \parbox[t]{3.4cm}{\raggedright SUVI 304~\AA, AIA 304~\AA} & \parbox[t]{7cm}{\raggedright Separation into an ascending upper segment and a partially draining lower segment.} \\
\parbox[t]{3.8cm}{\raggedright Post-eruption oscillation} & 14:17--16:37 & \parbox[t]{3.4cm}{\raggedright SUVI 304~\AA, FSI 304~\AA\ (comparison)} & \parbox[t]{7cm}{\raggedright Transverse oscillation of the upper segment continues after the bifurcation.} \\
\parbox[t]{3.8cm}{\raggedright Ascent into the outer corona} & 14:17 onward & \parbox[t]{3.4cm}{\raggedright SUVI 304~\AA, FSI 304~\AA} & \parbox[t]{7cm}{\raggedright Upper segment ascends as a coherent, detached structure that is tracked in SUVI ($\le 1.6\,R_\odot$) and FSI ($\le 6\,R_\odot$).} \\
\parbox[t]{3.8cm}{\raggedright Coronal mass ejection} & 18:24 & \parbox[t]{3.4cm}{\raggedright LASCO C2, FSI 304~\AA} & \parbox[t]{7cm}{\raggedright Appearance of a fully expanded CME front, spatially continuous with the FSI-tracked prominence material.} \\
\enddata
\end{deluxetable*}

\section{Analysis Methods} \label{sec:analysis}

\subsection{Identification of the Equilibrium and Rise Phase}

Prior to the onset of oscillations, the prominence remained in a quasi-static equilibrium state, characterized by a nearly constant height and stable intensity distribution. This phase represents the pre-disturbance configuration of the system before the arrival of an external perturbation that triggered the oscillations in the prominence.

\begin{figure}
\centering
\includegraphics[width=0.9\linewidth]{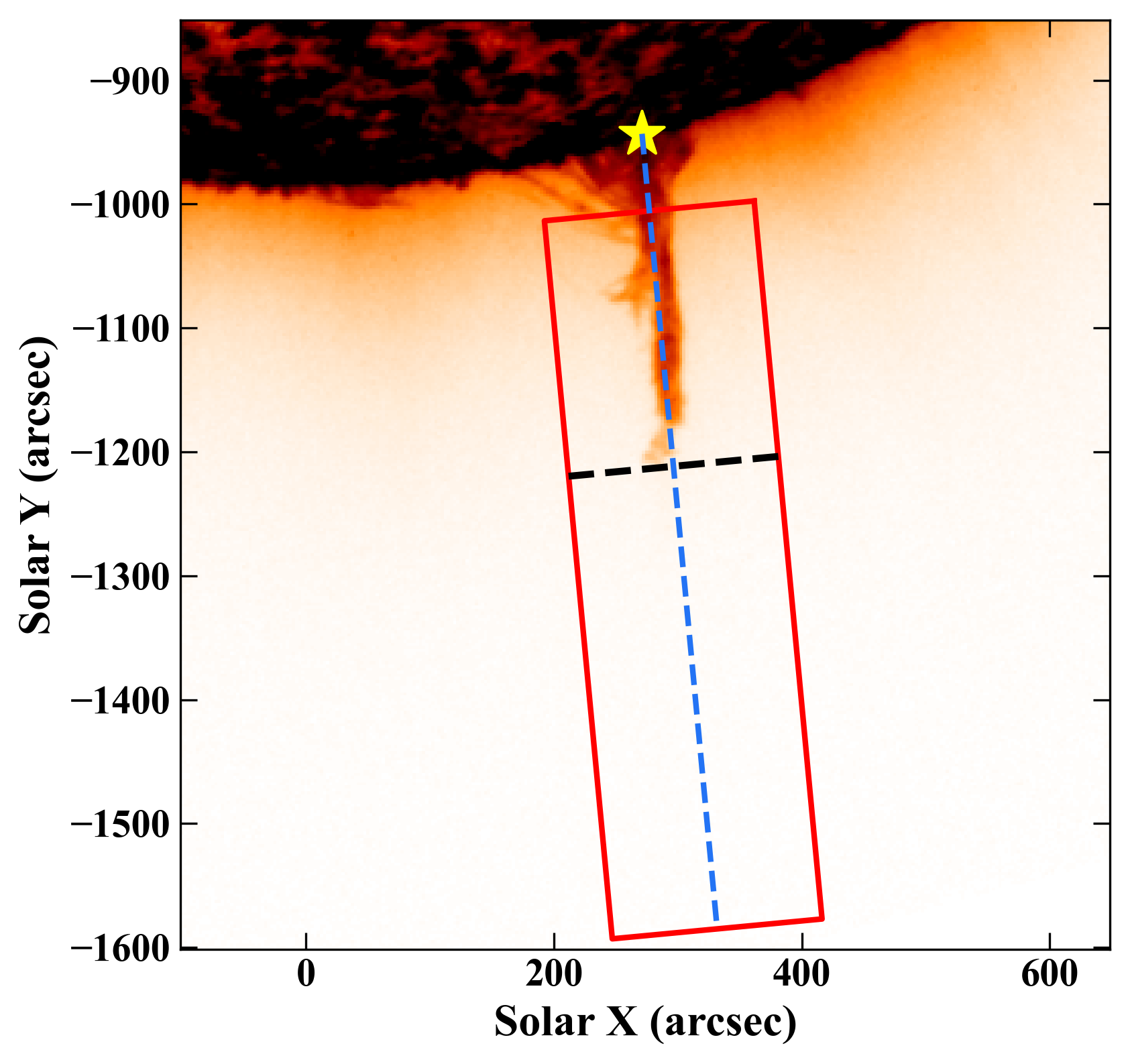}
\caption{SUVI 304 Å image (displayed in inverted color scale) of the prominence at 12:41:06 UT, showing the placement of the vertical slit along the prominence axis. The red box indicates the slit region, the blue dashed line represents the prominence axis, the yellow star marks the reference point near the limb (for height calculations), and the black dashed line denotes the equilibrium prominence height.}
\label{fig:slit}
\end{figure}

To track the large-scale vertical evolution of the prominence, a finite-width rectangular slit was defined by manually selecting two points along the prominence axis (blue dashed line in Figure~\ref{fig:slit}). The slit (red box in Figure~\ref{fig:slit}) was assigned a width of 124~Mm to ensure that horizontal displacements of the structure remained within the sampling region throughout the observation and a length of 400~Mm to capture the complete evolution of the prominence along the height.

\begin{figure}
\centering
\includegraphics[width=0.99\linewidth]{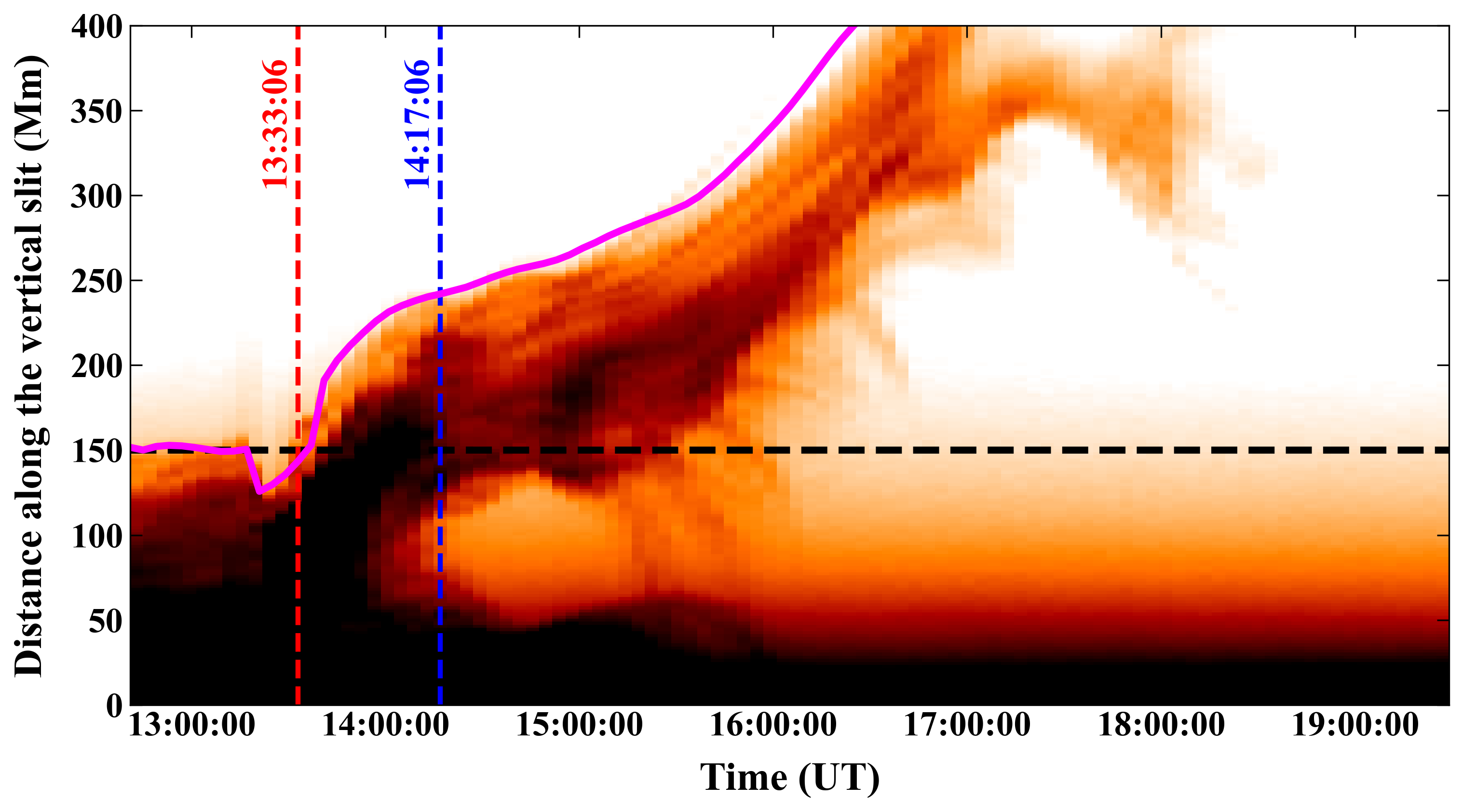}
\caption{
Time-distance (XT) (displayed in inverted color scale) map obtained from the vertical slit shown in Figure~\ref{fig:slit}. The horizontal dashed line indicates the equilibrium height ($\sim$150~Mm). The magenta curve represents the traced upper edge of the prominence. The red vertical dashed line marks the onset of the prominence rise, while the blue vertical dashed line denotes the time at which the prominence bifurcates.
}
\label{fig:xt_full}
\end{figure}

A time-distance (XT) map using SUVI 304~\AA\ was constructed by sampling the data in uniform increments of 1 pixel ($2.5''$ or $\sim 1.8$~Mm) along its length, which is the native pixel sampling of SUVI \citep{Darnel2022} and lies within the instrument’s spatial resolution. At each 1-pixel step, intensity values were extracted perpendicular to the slit axis across the full width using bilinear interpolation (order $=$ 1) and then averaged. Stacking the resulting profiles over time produced the $X$$T$ map (Figure~\ref{fig:xt_full}). From this map, the apparent equilibrium height of the prominence (top edge marked by a black dashed line in Figure~\ref{fig:slit}) is about 150 Mm from the base of the slit. The overall vertical extent of the prominence from its base near the limb (marked by a yellow star in Figure~\ref{fig:slit}) is about 195 Mm. A systematic upward displacement of the intensity ridge beyond this height marks the onset of the rise phase at approximately 13:33~UT.

\subsection{Extraction of Prominence Height-Time Profile}
\label{h-t profile}
To track the vertical motion of the prominence, the obtained XT map was initially smoothed with a 2D Gaussian filter ($\sigma = 1$; corresponds to 4 min $\times$ 1.8 Mm) to suppress high frequency noise for reliable edge detection. The upper edge of the prominence was then identified in each time frame by scanning from the top of the slit downward and locating the first coordinate where the magnitude of intensity gradient reached 10\% of the maximum gradient along that specific column. While the 10\% threshold may seem arbitrary, it was varied until the upper edge of the prominence was satisfactorily isolated for visual identification. Lower values detected background coronal noise, while higher values shifted the tracked boundary toward lower regions of the prominence. The resulting edge trajectory was subsequently smoothed using a Savitzky-Golay filter \citep{Savitzky1964} with a window length of 7 pixels and a polynomial order of 3, again to suppress high-frequency noise while preserving the large-scale evolution of the prominence height. This smoothed trajectory, overplotted as a magenta curve in Figure~\ref{fig:xt_full}, provides a continuous estimate of the prominence height as a function of time. 

\subsection{Construction of a Co-moving Slit for Transverse Oscillations}
\label{co-moving slit}

\begin{figure*}[t]
\begin{interactive}{animation}{Movies/Moving_slit.mp4}
\centering
\includegraphics[width=\textwidth]{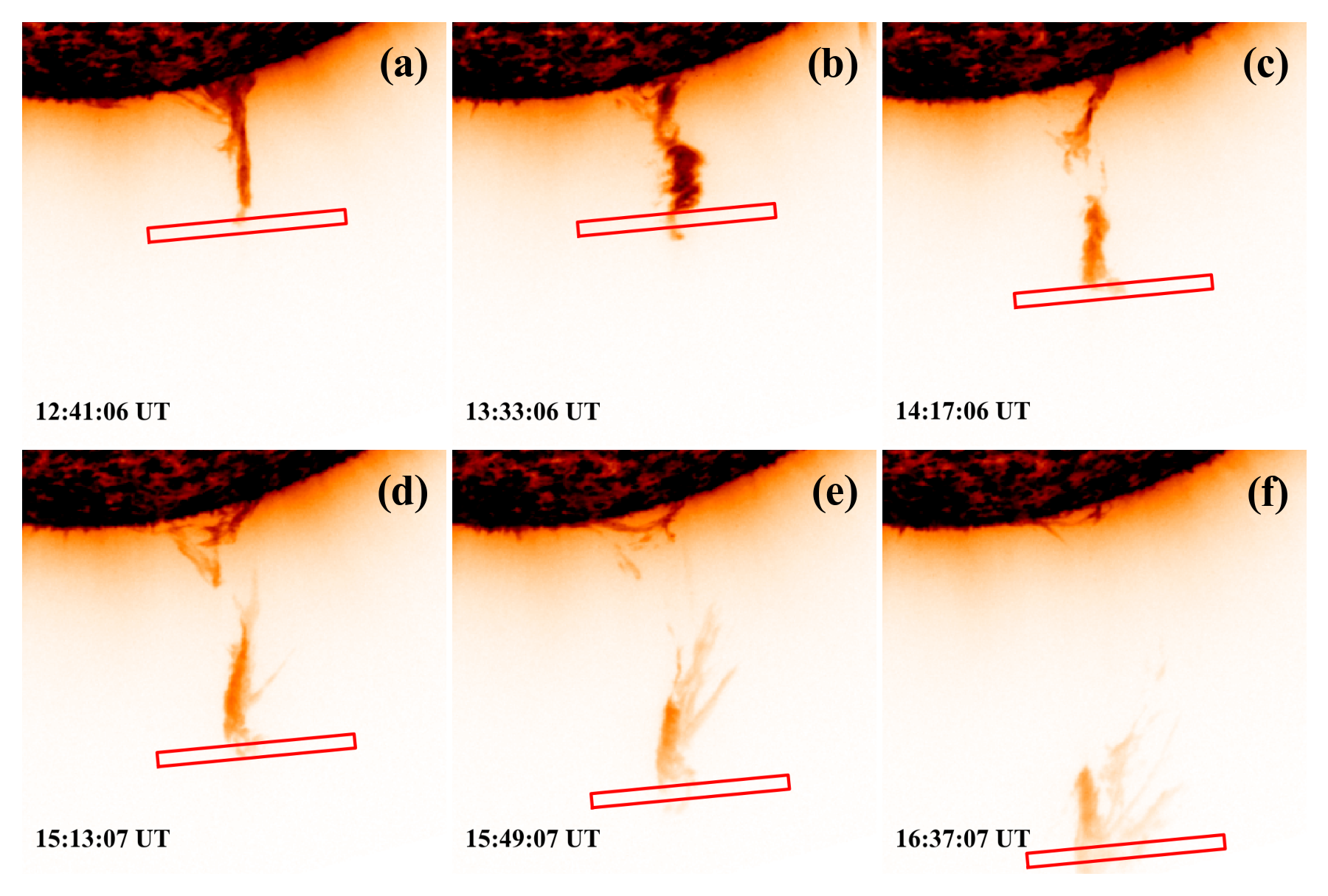}
\end{interactive}
\caption{
Evolution of the co-moving horizontal slit at different times (panels a-f), shown in SUVI 304 Å (displayed in inverted color scale). Panel (a) shows the prominence in its equilibrium state, panel (b) corresponds to the onset of the height rise, and panel (c) shows the bifurcation of the prominence into two components. Panels (d)-(f) illustrate the subsequent evolution of the co-moving slit as it follows the prominence motion, enabling consistent tracking of the same plasma segment. \\
\textit{This figure is available as an animation of SUVI 304 Å displayed in inverted color scale images. The animation starts on 2024 February 9 at 12:41~UT and ends at 16:37~UT, with a duration of 6~s.}
}
\label{fig:moving_slit}
\end{figure*}

The derived height-time profile was used to remove the large-scale vertical motion of the prominence and isolate its transverse oscillations. To achieve this, a horizontal slit oriented perpendicular to the prominence axis was defined (see Figure~\ref{fig:moving_slit}) . The slit had a width of 18 Mm and a length of 254 Mm. The slit length (oriented across the prominence axis) was chosen to capture the full transverse horizontal displacement throughout the observation, while the slit width (along the prominence axis) was used to average the pixel intensities and improve the signal-to-noise ratio.

The horizontal orientation of the slit allows the transverse displacement of the prominence to be directly captured along the slit direction. The slit was positioned near the upper end of the prominence, away from the solar limb, rather than at lower heights. \citet{Zhang2024} reported that the oscillation amplitude increases with height along the prominence, while decaying over time as the oscillation damps. Our analysis spans a longer duration, extending from the pre-eruptive slow-rise phase through bifurcation and into the post-eruption phase, and further excludes the initial forced-response interval driven by the EUV wave fronts (Section~\ref{sec:forced}). Over this extended interval, an oscillation tracked at lower heights would damp to an amplitude too small to measure reliably well before the oscillation cycles that continue through bifurcation and into the post-eruption phase could be captured. Positioning the slit near the upper end of the prominence therefore ensures a measurable transverse displacement signal across the full duration analyzed, including the oscillation cycles observed both before and after bifurcation.

To ensure that the same plasma segment was tracked throughout the observation, the slit was allowed to move vertically following the prominence motion. The displacement of the prominence apex was derived from the smoothed height-time profile (see magenta curve in Figure~\ref{fig:xt_full}), with the first frame taken as the reference for all subsequent measurements. The central position of the slit at each time step was then adjusted according to this displacement, effectively constructing a co-moving slit that compensates for the bulk vertical motion of the prominence, thereby ensuring that the same plasma segment is consistently tracked throughout the observation.

The evolution of the prominence along with the co-moving slit marked at different times is shown in Figure~\ref{fig:moving_slit} (see also the accompanying animation).

\subsection{Interaction between the EUV wave fronts and the prominence} \label{sec:forced}

\citet{Zhang2024} identify two successive EUV wave fronts, WF1 and WF2, driven respectively by the associated CME and a coronal jet, arriving at the prominence at $\approx$13:09~UT and $\approx$13:17~UT. To derive the propagation speeds of these wave fronts, they use a curved slice oriented along the wave-front geometry in AIA 193~\AA. Separately, to characterize the prominence oscillation itself, they use ten straight slices perpendicular to the prominence axis in 304~\AA, fitting the transverse displacement from the arrival time of WF1 ($t_0 \approx 13{:}09$~UT) onward.

\begin{figure*}
\centering
\begin{minipage}[t]{0.47\textwidth}
    \centering
    \includegraphics[width=\linewidth]{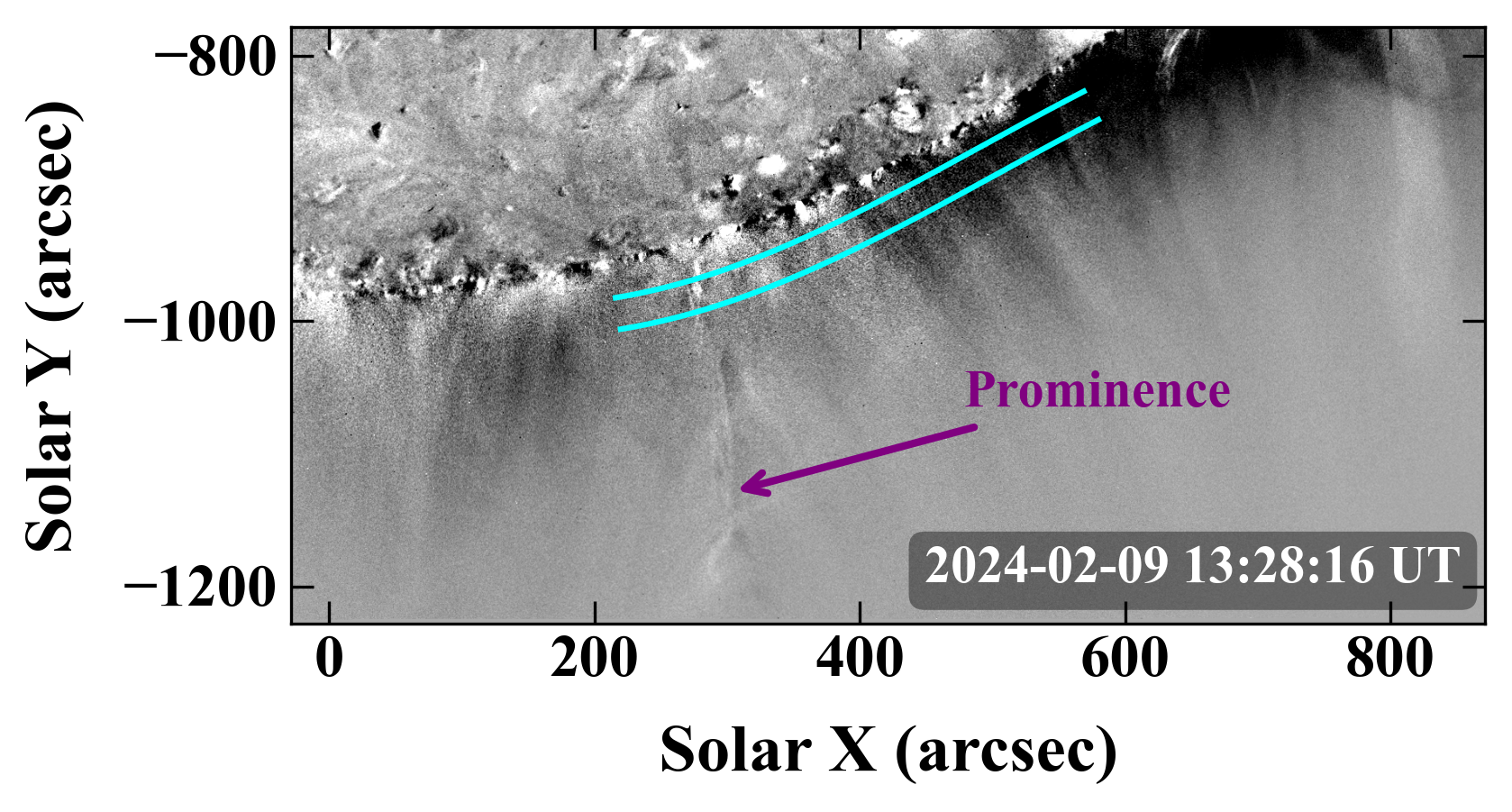}
    \par\smallskip
    (a)
\end{minipage}
\hfill
\begin{minipage}[t]{0.50\textwidth}
    \centering
    \includegraphics[width=\linewidth]{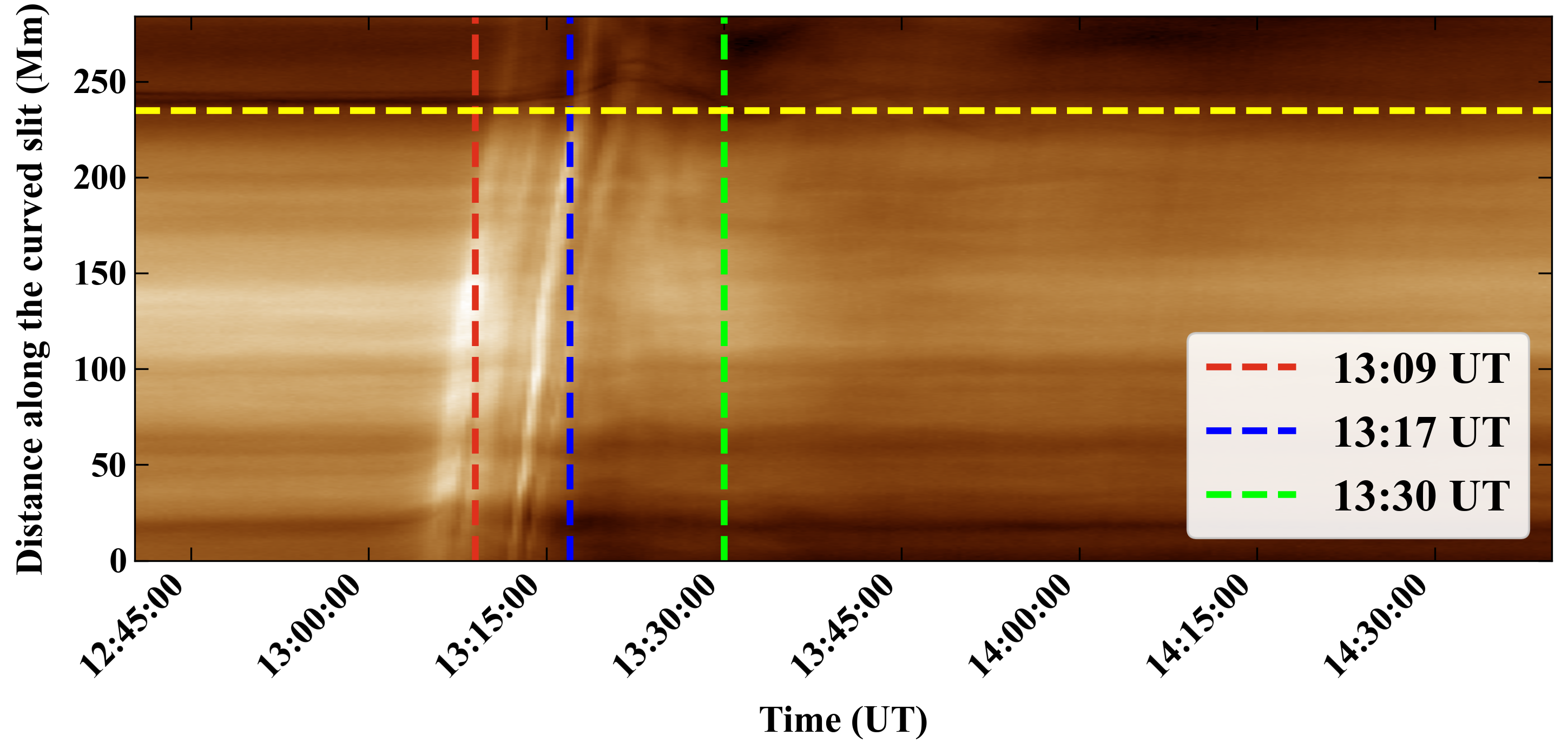}
    \par\smallskip
    (b)
\end{minipage}

\vspace{0.5cm} 

\begin{minipage}[t]{0.48\textwidth} 
    \centering
    \includegraphics[width=\linewidth]{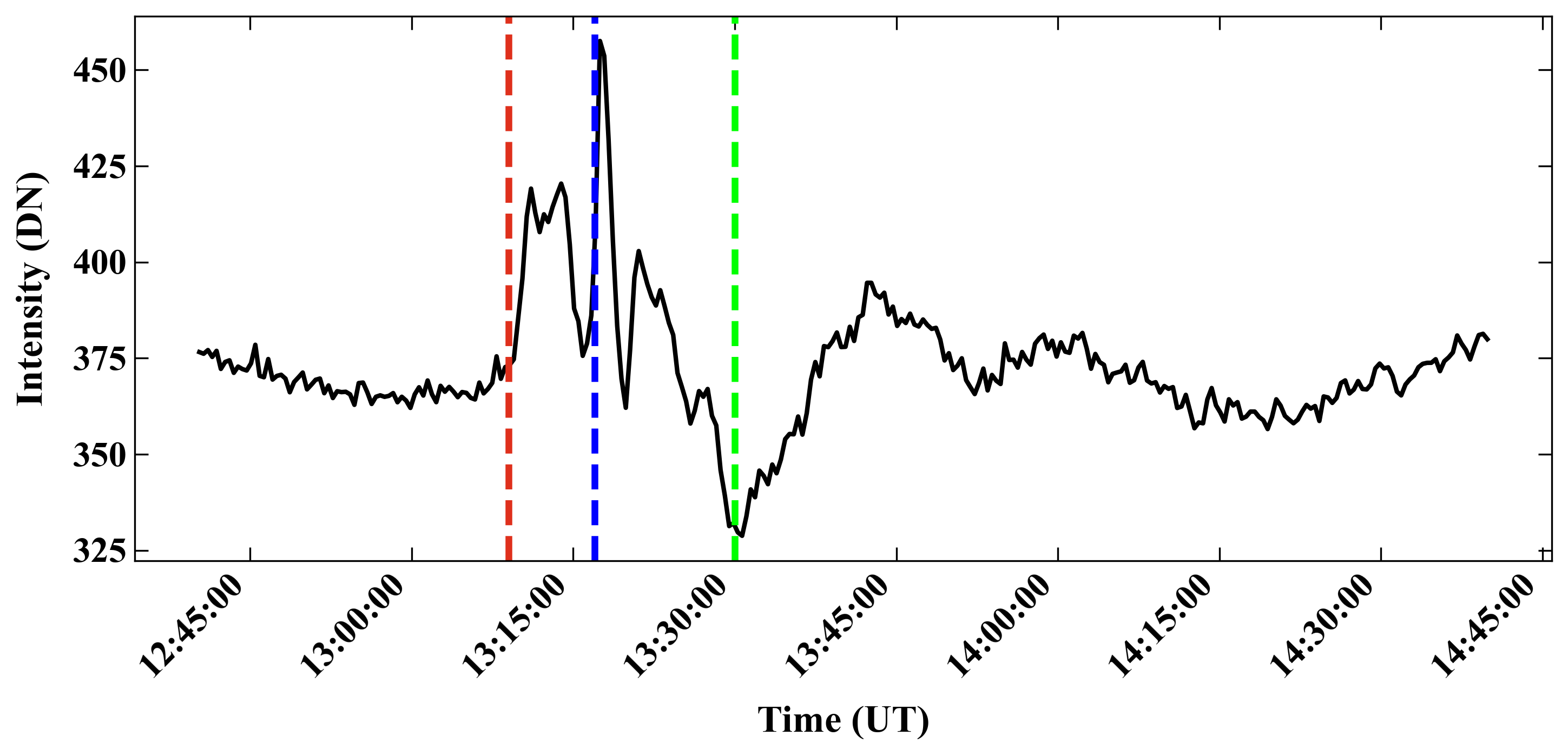}
    \par\smallskip
    (c)
\end{minipage}

\caption{(a) AIA 193~\AA\ base-difference image showing the placement of the curved slit. The two cyan lines mark the edges of the 17~Mm wide slit, across which intensities are averaged to construct the time-distance map. The slit is oriented along the wave-front geometry. (b) Time-distance map along the slit shown in panel (a), constructed from AIA 193~\AA\ images. The yellow dashed line marks the fixed distance (235~Mm) at which the lightcurve in panel (c) is extracted. (c) Lightcurve extracted along the slit at 235~Mm from the slit origin (western end of the slit), showing the intensity variation with time at this location. The red, blue and green dashed lines in panels b) and c) mark the arrival of WF1 (13:09~UT), WF2 (13:17~UT), and the subsequent local minimum in intensity ($\approx$13:30~UT) marking the end of the forced response, respectively.}
\label{fig:forcedwave}
\end{figure*}

Here, our aim is to isolate the transverse oscillation of the prominence from its directly forced response to the passage of the wave fronts. For this, we require a complementary piece of information: the time at which the prominence's motion decouples from the wave forcing and evolves under its own restoring force alone. To determine this, we adapt a curved slit oriented along the wave-front geometry (Figure~\ref{fig:forcedwave}(a)), following the approach of \citet{Zhang2024}. We tested this slit in the AIA 171, 193, and 211~\AA\ channels. Both
193~\AA\ (Fe~\textsc{xii}, $\sim$1.5~MK) and 211~\AA\ (Fe~\textsc{xiv}, $\sim$2~MK) show clear contrast for the wave front and the prominence response, whereas 171~\AA\ (Fe~\textsc{ix}, $\sim$0.6~MK) was comparatively noisier. We construct a time-distance map using a curved slit (length 284 Mm, width 17 Mm) in the AIA 193~\AA\ channel (Figure~\ref{fig:forcedwave}b). To characterize the intensity variation associated with the wave passage more directly, we additionally extract a lightcurve (shown in Figure~\ref{fig:forcedwave}c) along the slit at a fixed distance of 235~Mm from the slit origin (on the western side of the slit) as marked by the yellow dashed line in Figure~\ref{fig:forcedwave}b.

The time-distance map (Figure~\ref{fig:forcedwave}b) shows that, following the arrival of WF1 at 13:09~UT and WF2 at 13:17~UT, the prominence continues to display a coherent, wave-correlated displacement signature for several additional minutes, rather than an immediate return to an undisturbed oscillatory pattern. The lightcurve extracted at 235~Mm along the slit (Figure~\ref{fig:forcedwave}c) makes this behavior directly quantifiable: rather than a single isolated impulse, the intensity exhibits a series of enhancements between the arrival of WF1 and $\approx$13:30~UT, before settling into a pronounced dip that marks the transition to the prominence's natural oscillation. This multi-peaked structure indicates that the two wave fronts do not act as independent, instantaneous impulses on the prominence, but rather produce a single compound forced response, during the passage of the EUV waves across the prominence body, consistent with the finite spatial width of the wave fronts themselves.

The curved slit, tracing the wave-front geometry in AIA 193~\AA\ (Figure~\ref{fig:forcedwave}b,c), is positioned near the solar limb, whereas the co-moving slit, tracking the oscillating parcel in SUVI 304~\AA\ (Figure~\ref{fig:xt_comparison}a), samples the prominence near its upper extent. Since the wave-front propagates outward from its origin and sweeps obliquely across the prominence (see Figure 6 of \citealt{Zhang2024}), it does not make contact with the full length of the prominence simultaneously. As a result, the arrival of the wave-front, and hence the timing of the peak wave disturbance in AIA and in SUVI, slightly differs between the two slit locations. In contrast, the end of the forced response coincides across both diagnostics within the observational cadence, as the intensity minimum in AIA 193~\AA\ and the first transverse-displacement minimum in SUVI 304~\AA\ both occur at $\approx$13:30~UT.

Following this result, we do not treat the full rise-and-fall following the wave arrivals as part of the natural oscillation. Instead, we identify the first local minimum in the tracked position-time series following the peak displacement (Section~\ref{sec:fitting}) as the point at which the directly forced response gives way to the prominence's own decaying oscillation, and restrict the fit accordingly.

\subsection{Time-Distance Analysis and Oscillation Fitting} \label{sec:fitting}

\begin{figure*}
\centering
\begin{minipage}[t]{0.48\textwidth}
    \centering
    \includegraphics[width=\linewidth]{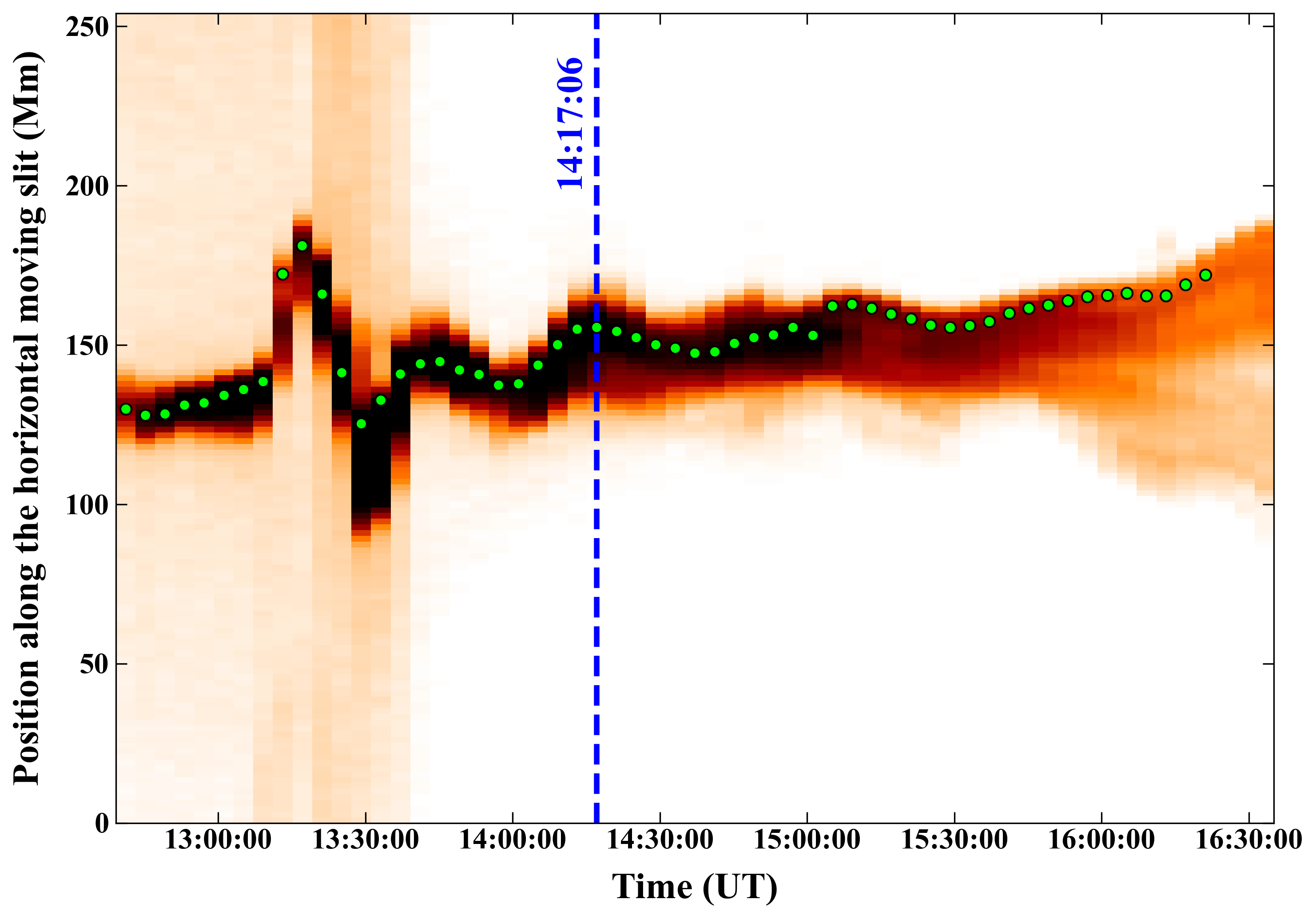}
    \par\smallskip
    (a)
\end{minipage}
\hfill
\begin{minipage}[t]{0.48\textwidth}
    \centering
    \includegraphics[width=\linewidth]{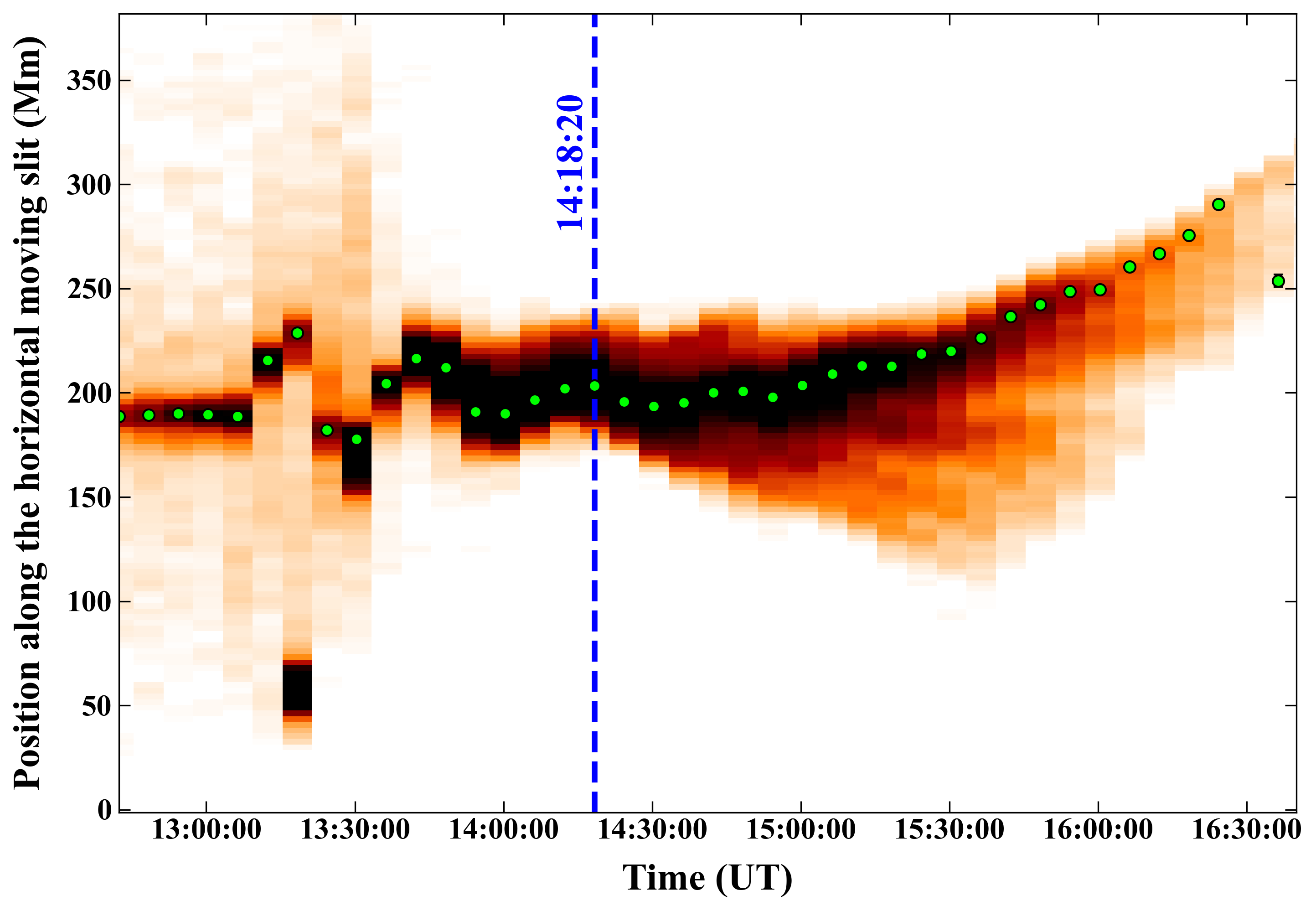}
    \par\smallskip
    (b)
\end{minipage}
\caption{Time-distance map (displayed in inverted color scale) from the co-moving horizontal slit shown in Figure~\ref{fig:moving_slit}. Green dots represent Gaussian centroid positions with respective uncertainties. The vertical dashed line marks the prominence bifurcation time. (a) SUVI 304~\AA. (b) FSI 304~\AA.}
\label{fig:xt_comparison}
\end{figure*}

The XT map obtained from the co-moving horizontal slit reveals clear transverse oscillations that decay over time, as shown in
Figure~\ref{fig:xt_comparison}(a).

At each time step, the prominence position along the slit was determined by fitting a Gaussian function to the intensity profile, yielding the centroid position and its associated uncertainty. The temporal evolution of these centroid positions (marked as green dots in Figure~\ref{fig:xt_comparison}) forms the position-time series used to characterize the oscillatory motion.

The oscillatory behavior in the pre-eruption and post-eruption phases is segregated based on the time at which the prominence separates into two segments, as identified through visual inspection of the time sequenced SUVI and AIA 304~\AA\ observations, where the structure clearly separates into two distinct components.

A damped sinusoidal function with a linear trend was then fitted to this position-time series in each segment to model the transverse oscillations, using the \texttt{lmfit} package \citep{Newville2014} in Python, which performs Levenberg-Marquardt minimization. The functional form of the model is given by
\begin{equation}
y(t) = A \sin\left(\frac{2\pi}{P}t + \phi \right) e^{-t/\tau} + y_0 + k t,
\end{equation}
where $A$ is the amplitude, $P$ is the oscillation period, $\phi$ is the initial phase, $\tau$ is the damping time, $y_0$ is the initial position of the prominence, and $k$ represents the linear horizontal drift of the prominence.

The velocity amplitude of the oscillation is given by
\begin{equation}
v = \frac{2\pi A}{P},
\end{equation}
with the corresponding uncertainty propagated from the uncertainties in $A$ and $P$ (the $1\sigma$ standard errors returned by \texttt{lmfit} from the covariance matrix of the fit) as
\begin{equation}
\sigma_{v} = \sqrt{\left(\frac{2\pi}{P}\,\sigma_A\right)^2 + \left(\frac{2\pi A}{P^2}\,\sigma_P\right)^2}.
\end{equation}

\begin{figure*}[!t]
\centering
\includegraphics[width=0.9\textwidth]{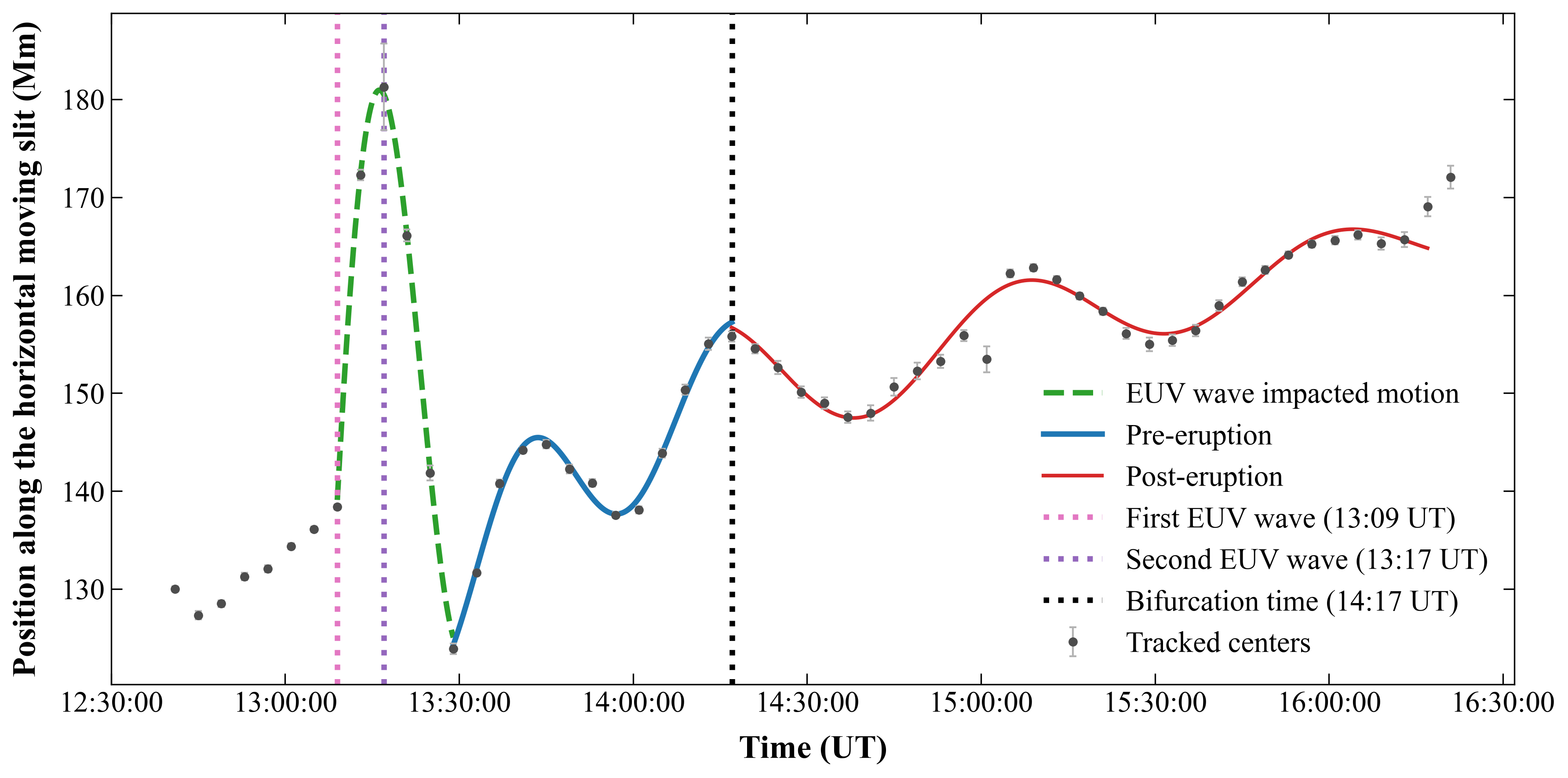}
\caption{Position of the tracked prominence centers along the horizontal moving slit (SUVI) as a function of time. Blue and red solid lines represents the damped sinusoidal fitting for pre-eruption phase and post-eruption phase.}
\label{fig:oscillation_fit}
\end{figure*}

Guided by the discussion in Section~\ref{sec:forced}, we exclude the directly forced response to the two EUV wave fronts from the fit in the pre-bifurcation phase. We take the first minimum as $t_0$ for the pre-bifurcation fit, marking the onset of the natural transverse oscillation. The interval prior to this point reflects the compound forced response to the two wave fronts rather than the prominence's intrinsic oscillatory motion, and is excluded from the quantitative comparison in Section~\ref{sec:results}. The post-bifurcation phase is fit independently, beginning at the bifurcation time (14:17:06 UT). The tracked positions and all three regimes are shown in Figure~\ref{fig:oscillation_fit}.

\subsection{Modeling the Temporal Evolution of the Prominence}\label{subsec:model}

\begin{figure*}
\begin{interactive}{animation}{Movies/magnetic_field.mp4}
\centering
\includegraphics[width=\linewidth]{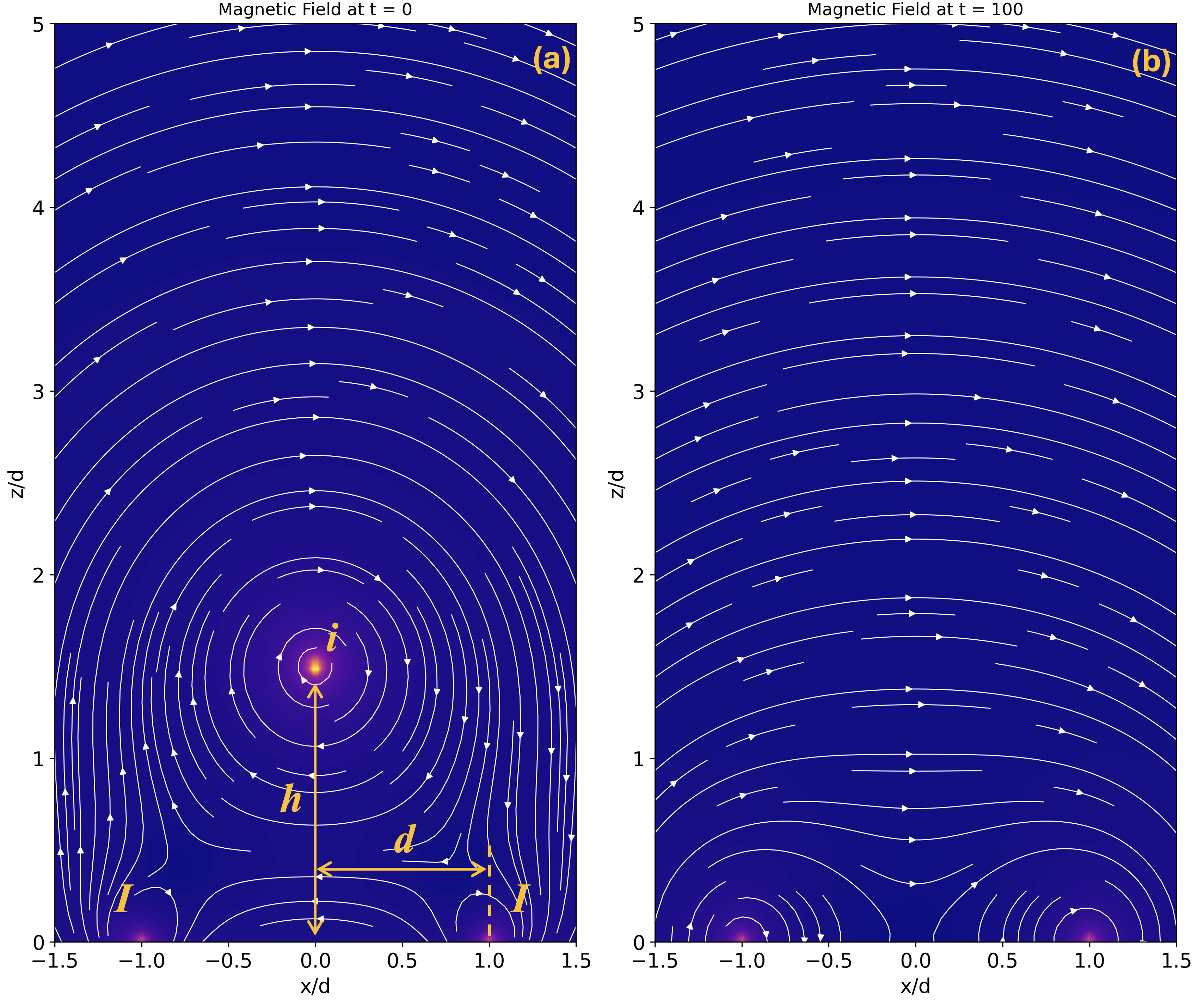}
\end{interactive}
\caption{
Schematic representation of the inverse-polarity prominence configuration, adapted from \citet{Dmitrii2018}. The prominence is modeled as a line current $i$ suspended at a height $h$ above the surface, within a magnetic dip formed by two sub-photospheric line currents $I$ located at $x/d=\pm1$. Panels (a) and (b) show the magnetic configuration at $t=0$ and at $t = 100$ (after the prominence has erupted and left the plotted domain, revealing the magnetic dip generated by the sub-photospheric currents), respectively.
\\ \textit{This figure is available as a 10~s animation illustrating the temporal evolution of the model.} }
\label{fig:theory_snapshot}
\end{figure*}

To qualitatively interpret the observed evolution of the prominence oscillations, we employ the theoretical framework developed by \citet{Dmitrii2018}, which relates the oscillatory properties of a prominence to its underlying magnetic configuration. However, since our prominence is eruptive, in contrast to their model, we consider an unstable inverse-polarity configuration here. A schematic of our model configuration is shown in Figure~\ref{fig:theory_snapshot} (see also the accompanying animation). In this model (Fig.~\ref{fig:theory_snapshot}a), the prominence is represented as a massive line current $i$ suspended at height $h$ above the photosphere, in a magnetic dip generated by two sub-photospheric line currents of strength $I$, separated by a distance $2d$. The photospheric currents alone form a dip. In Fig.~\ref{fig:theory_snapshot}a, this dip is masked by the field of the prominence current and its mirror current, which dominates the total field in this region; the photospheric-current dip becomes apparent in Fig.~\ref{fig:theory_snapshot}b, once the prominence has erupted and left the plotted domain. Note, for $I/i$ being about 1 or above, this inverse-polarity model becomes vertically unstable for $h/d>1$.

In this model, the prominence current flows into the plane of the figure ($i<0$). While in this configuration too, the horizontal equilibrium of the prominence is ensured by the left-right symmetry of the photospheric currents, and the vertical equilibrium is set by the balance between the Lorentz forces exerted by the two photospheric currents (see the first term in Eq.~(\ref{eq:equilibrium})) and the prominence's own image (`mirror') current induced in the photosphere below it (see the second term in Eq.~(\ref{eq:equilibrium})), and the resultant vertical force $\mathcal{F}_\mathrm{vert}$ arising from other physical mechanisms such as gravity, magnetic tension, hoop force, etc.
The vertical force balance can therefore be expressed as
\begin{equation}
\frac{2k_1 h}{d^2+h^2} + \frac{k_2}{2h} + \mathcal{F}_\mathrm{vert} = 0,
\label{eq:equilibrium}
\end{equation}
with magnetic constants defined in terms of $k_1$ and $k_2$ as 
\begin{equation}
k_1 = \frac{\mu_0 I i}{2\pi}, \qquad k_2 = \frac{\mu_0 i^2}{2\pi},
\label{eq:k1k2}
\end{equation}
where, in the inverse-polarity configuration adopted here, the overarching external magnetic field pulls downward on the prominence. Note that in the original framework by \citet{Dmitrii2018}, $k_1$ was defined using the absolute magnitudes of $I$ and $i$ to represent a normal-polarity configuration with an upward external Lorentz force. To adapt this same mathematical framework for an inverse-polarity configuration, we adopt the convention $k_1 < 0$ throughout this work, with $\mu_0$ being the vacuum permeability. All forces in this model are similarly normalized per unit length along the current direction.

When the prominence is perturbed by an external disturbance, it undergoes horizontal and vertical displacements $(x,z)$ from its equilibrium position. Its dynamics are governed by
\begin{equation}
\mathcal{R}\frac{d^2 x}{dt^2} = F_x - \delta\frac{dx}{dt}, \qquad
\mathcal{R}\frac{d^2 z}{dt^2} = F_z - \delta\frac{dz}{dt},
\label{eq:motion}
\end{equation}
where $F_x$ and $F_z$ are the horizontal and vertical components, respectively, of the net force acting on the prominence current, and $\mathcal{R}$ is the linear mass density of the prominence.
The dissipative terms on the right-hand sides of Equations~(\ref{eq:motion}) with $\delta=0.05$ are added to mimic effective damping of the prominence dynamics \citep[see e.g.][]{2018JASTP.179..149Z}.
As derived by \citet{Dmitrii2018}, these forces are given by:

\begin{equation}
F_x = \frac{2k_1 x\left[(h+z)^2+x^2-d^2\right]}{D},
\label{eq:Fx}
\end{equation}
\begin{equation}
F_z = \frac{2k_1(h+z)\left[d^2+x^2+(h+z)^2\right]}{D} 
      + \frac{k_2}{2h+z} + \mathcal{F}_\mathrm{vert},
\label{eq:Fz}
\end{equation}
where
\begin{equation*}
D = (d^2-x^2)^2 + 2(d^2+x^2)(h+z)^2 + (h+z)^4,
\end{equation*}
and $\mathcal{F}_\mathrm{vert}$ is prescribed by the equilibrium condition given by Eq.~(\ref{eq:equilibrium}) and kept constant during the evolution.

In Equations~(\ref{eq:Fx})-(\ref{eq:Fz}), $x$ and $z$ represent the instantaneous horizontal and vertical displacements, respectively, of the prominence from its equilibrium position, where the $z$-axis is directed upward. To isolate the horizontal oscillations and derive their oscillation properties, we restrict our analysis to small-amplitude horizontal perturbations ($x \ll d$). By performing a Taylor expansion of the horizontal restoring force $F_x(x, z)$ with respect to $x$ and linearizing the expression, we obtain the first-order terms in $x$. The denominator simplifies directly to $D(0, z) = \left(d^2 + (h+z)^2\right)^2$, yielding the linearized horizontal restoring force \citep[see also][]{2016A&A...590A.120K}
\begin{equation}
F_x \approx \left[ \frac{2k_1 \left((h+z)^2 - d^2\right)}{\left(d^2 + (h+z)^2\right)^2} \right] x.
\end{equation}

Substituting this linearized force into the horizontal equation of motion ($\mathcal{R}\ddot{x} = F_x$) yields the harmonic oscillator equation $\ddot{x} + \omega_x^2 x = 0$, where the height-dependent horizontal oscillation frequency is given by
\begin{equation}
\omega_x^2 = -\frac{2k_1\left((h+z)^2-d^2\right)}{\mathcal{R}\left(d^2+(h+z)^2\right)^2}.
\label{eq:omega1}
\end{equation}

Note that in our adopted inverse-polarity configuration, $k_1 < 0$. In fact, we initialize the model with $h/d=1.5$, $k_1=-0.5$, and $k_2=1$ in normalized units.
These values are not derived from the present observations; they are chosen to represent a typical inverse-polarity configuration, since the model is intended as a qualitative illustration of the observed behavior. For a typical prominence with electric current
$i=2\times10^{10}$\,A, volume mass density $10^{-10}$\,kg\,m$^{-3}$, radius 5\,Mm, and the half-distance between the photospheric currents $d=10$\,Mm,  the characteristic time scale is $\tau_0 = \sqrt{\mathcal{R}d^2/k_2} \approx 1.65$\,min.
Since the dimensional parameters enter the model only through this time scale, different combinations of the prominence density, radius, current and the half-distance between the photospheric currents giving the same $\tau_0$ should result in identical dynamics.
This configuration results in $h+z > d$, which, in turn, yields $\omega_x^2 > 0$. This implies that the motion of the prominence is oscillatory in the horizontal direction. However, a displacement in the vertical direction results in a net force upwards and, consequently, the prominence continues to rise along a trajectory determined by Equation~(\ref{eq:Fz}).
Linearizing Equation~(\ref{eq:Fz}) about the equilibrium gives $\omega_z^2 \approx -0.007$, compared with $\omega_x^2 \approx 0.118$ from Equation (10) with $z \to 0$ (in units of $\tau_0^{-2}$). The equilibrium is therefore weakly unstable in the vertical direction, with perturbations growing as $e^{\gamma t}$ at the rate $\gamma = \sqrt{-\omega_z^2} \approx 0.085$, about four times smaller than the horizontal oscillation frequency $\omega_x \approx 0.34$. This weak instability, acting on a small initial vertical perturbation, allows the prominence to perform several horizontal oscillation cycles during the slow rise before erupting.

\begin{figure*}
\centering
\includegraphics[width=0.299\textwidth]{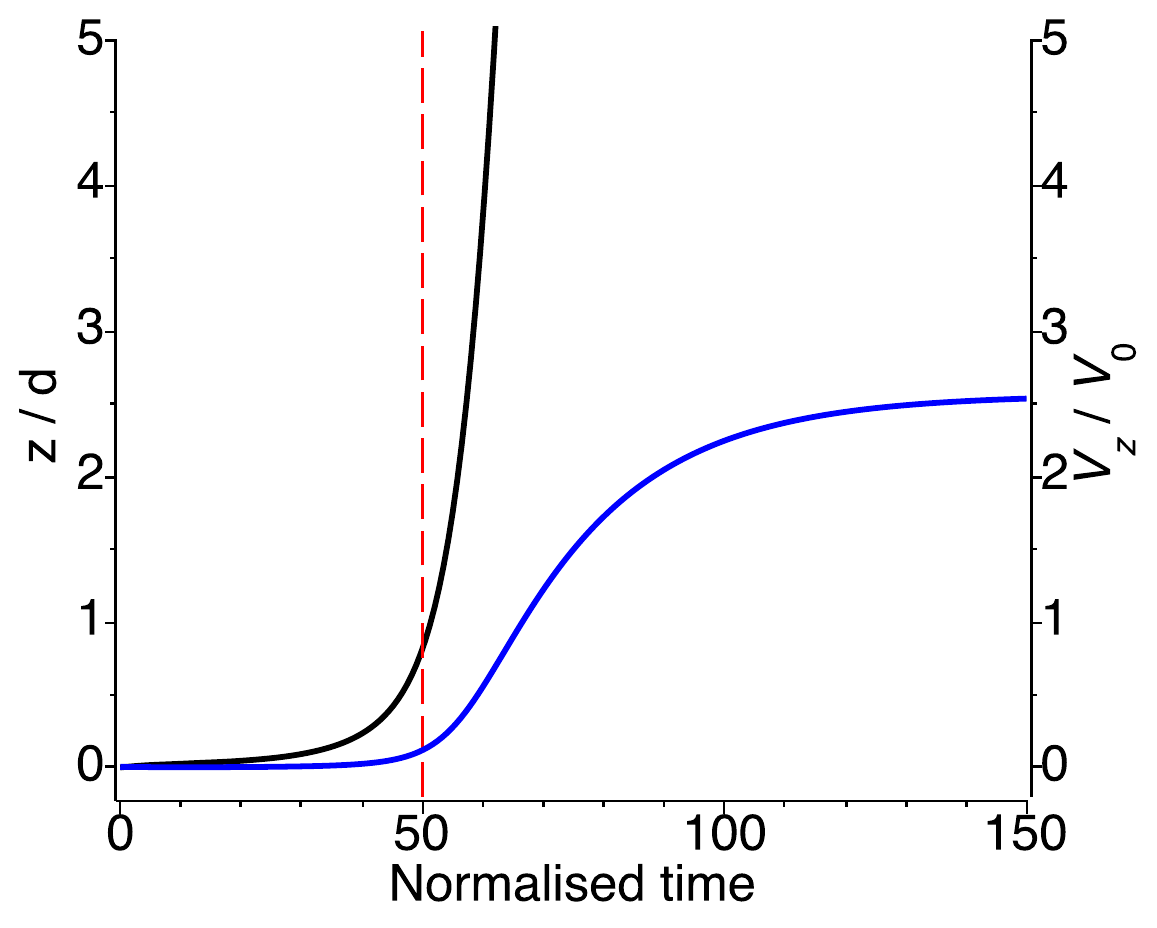}
\includegraphics[width=0.42\textwidth]{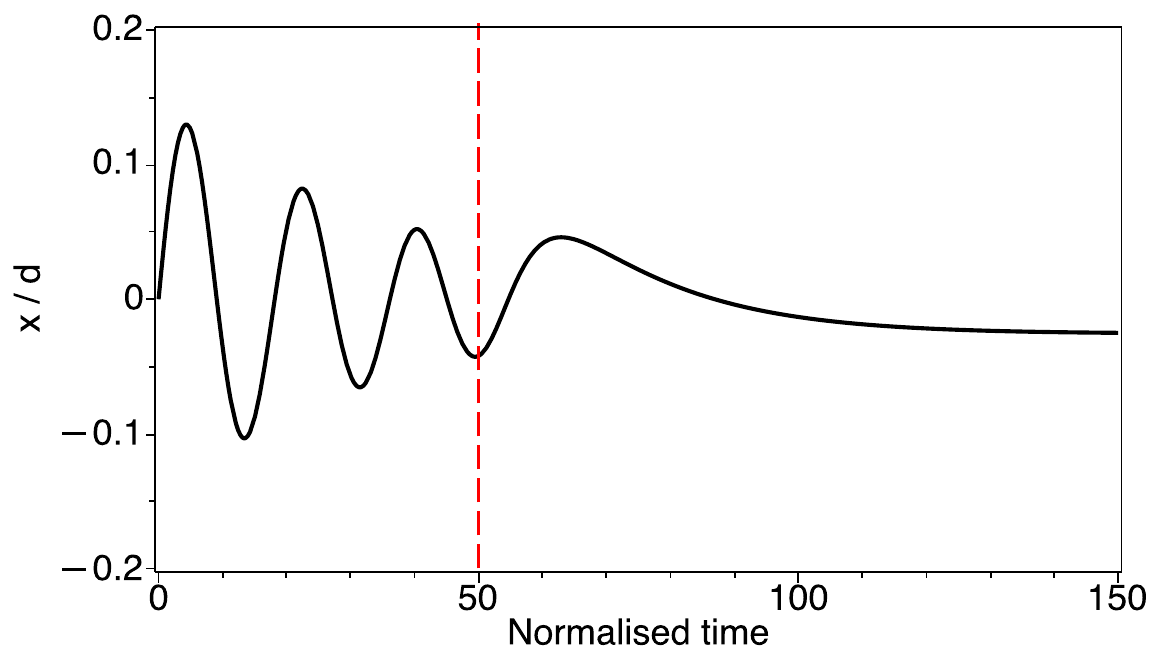}
\includegraphics[width=0.27\textwidth]{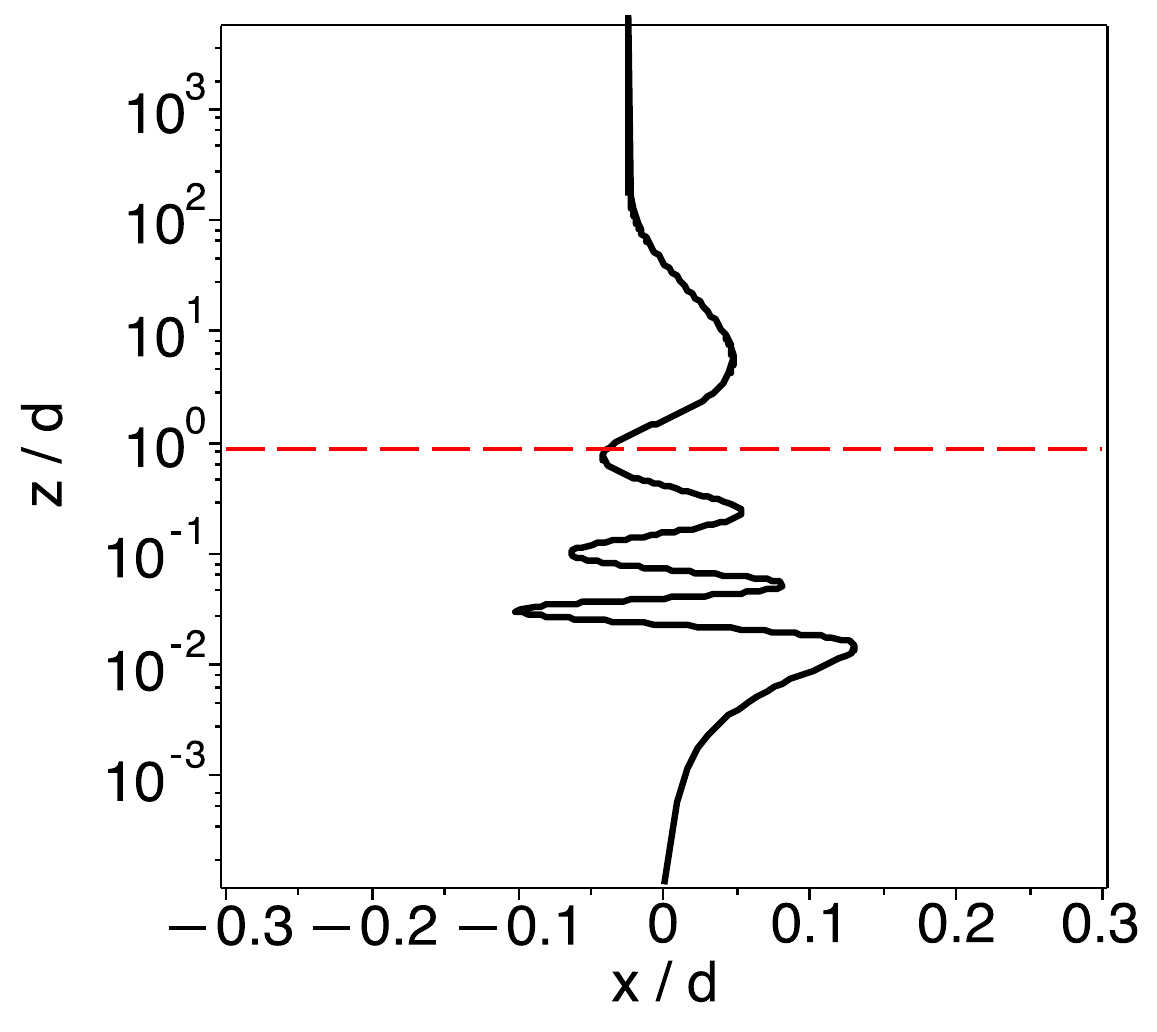}
\caption{Left: temporal evolution of the normalized prominence height ($z/d$, black) and vertical velocity ($V_z/V_0$, blue). Middle: temporal evolution of the normalized horizontal displacement ($x/d$), showing progressively lengthening oscillation cycles. The red dashed line marks the onset of the rapid acceleration phase ($t \approx 50$), corresponding to the eruption of the prominence. Right: full prominence trajectory. The horizontal dashed line shows approximate prominence position at $t=50$. The model time is normalized to $\sqrt{\mathcal{R}d^2/k_2}$, and the vertical velocity $V_z$ is normalized to $V_0 = \sqrt{k_2/\mathcal{R}}$.
}
\label{fig:theory_nonlinear}
\end{figure*}

The full spatiotemporal dynamics $x(t)$ and $z(t)$ of the prominence in the horizontal and vertical directions is illustrated in Figure~\ref{fig:theory_nonlinear}, which is obtained by double integrating the full nonlinear equations of motion (Equations~\ref{eq:Fx}-\ref{eq:Fz}), performed numerically with \emph{dsolve} function in the mathematical environment \emph{Maple} v.2025.2, using $x(0)=0$, $z(0)=0$, $\dot{x}(0)=0.05$, and $\dot{z}(0)=0.0041$. Such an initial condition corresponds to the excitation of the prominence by a predominantly horizontal external impact. The left panel additionally shows normalized prominence velocity in the vertical direction (blue curve) as obtained from the derivative of the corresponding displacement. As can be seen, the vertical evolution of the prominence highlights an initial slow-rise phase, which later transitions into a rapid vertical acceleration. Based on the onset of a sharp change in vertical velocity, we determine a characteristic time $\tau \approx 50$ (in normalized units), after which we consider the prominence as erupted.

The concurrent horizontal kinematics (middle panel, $x/d$) of the prominence indicate an oscillatory motion with a frequency that is directly dictated by its vertical evolution. Indeed, as may be inferred from Equation~(\ref{eq:omega1}), with parameters $k_1$, $h$, $d$, and $\mathcal{R}$ constant, the frequency $\omega_x$ depends inversely on the vertical displacement $z$. Further, this implies that during the slow-rise phase, when $z$ is varying slowly, the change in the oscillation frequency (or period) is negligible. Once the rapid acceleration begins post eruption, the sharp change in $z$ results in a significantly lower frequency (longer period).
Using the normalisation time scale $\sqrt{\mathcal{R}d^2/k_2} \approx 1.65$\,min, we estimate the horizontal oscillation period to vary approximately from $\sim30$\,min before the eruption to $\sim65$\,min during and after the eruption.
Note that the oscillation period continues to grow as $z$ keeps on increasing, and at larger heights ($z \gg d$) Equation~(\ref{eq:omega1}) can be further simplified to   
\begin{equation}
\omega_x^2 \approx -\frac{2k_1}{\mathcal{R} z^2}.
\label{eq:12}
\end{equation}
In this regime, the horizontal oscillation frequency scales linearly as $\omega_x \propto z^{-1}$, producing an oscillation period $P_x(z) \propto z$. The full prominence trajectory $z(x)$ is shown in the right-hand panel of Figure~\ref{fig:theory_nonlinear}.
 
\section{Results} \label{sec:results}

\subsection{Comparison of Oscillation Parameters Before and After Eruption}
\label{subsec:before_after}

A clear and systematic change in the oscillation properties is observed across the bifurcation of the prominence, as shown in Figure~\ref{fig:oscillation_fit} and Table~\ref{tab:table1}. The oscillation period $P$ increases from $34.3 \pm 0.5 $ minutes in the pre-eruption phase to $54.3 \pm 0.4 $ minutes in the post-eruption phase. The increase in oscillation period is accompanied by a decrease in amplitude $A$, from $9.5 \pm 0.4 $~Mm before the eruption to $6.9 \pm 0.4$~Mm after eruption. This reduction in amplitude indicates continued damping of the oscillatory motion, with a corresponding increase in damping time, $\tau$ from $76.8 \pm 8.4 $ to $133.6 \pm 17.7 $ minutes. The characteristic velocity amplitude $v$, estimated as $v \sim 2\pi A/P$, correspondingly decreases from $29.0 \pm 1.2$ to $13.2 \pm 0.7~\mathrm{km\,s^{-1}}$ in the post-eruption phase, due to the combined effect of increasing period and decreasing amplitude. Note that the estimated velocities and amplitudes are subject to projection effects, whereas the periods and damping times are not affected. The change in individual parameters is examined in terms of the underlying magnetic support in Section ~\ref{sec:discussion}.

Observational evidence of evolving oscillation periods has been reported in earlier studies of prominence oscillations. For instance, \citet{Isobe2006} observed an increasing oscillation period during the destabilization phase of a filament, interpreting this as an indication of a weakening restoring force. Furthermore, \citet{Bi2014} reported variations in oscillation periods during successive cycles. In that study, the oscillations were longitudinal and governed by gravity projected along dipped magnetic field lines. Although the physical mechanisms in these observational studies differ from the transverse oscillations analyzed here, both cases exhibit a similar trend of an increasing period during the dynamic evolution of the prominence, suggesting a possible weakening of the effective restoring force.

\begin{table}
\centering
\caption{Comparison of oscillation parameters before and after the prominence eruption.}
\label{tab:table1}
\begin{tabular*}{\columnwidth}{@{\extracolsep{\fill}}lcc}
\tableline
Parameter & Pre-Eruption & Post-Eruption \\
\tableline
Amplitude, A (Mm) &
$9.5 \pm 0.4$ &
$6.9 \pm 0.4$ \\

Period, P (min) &
$34.3 \pm 0.5$ &
$54.3 \pm 0.4$ \\

Damping Time, $\tau$ (min) &
$76.8 \pm 8.4$ &
$133.6 \pm 17.7$ \\

Velocity Amplitude, $v$  ($\mathrm{km\,s^{-1}}$) &
$29.0 \pm 1.2$ &
$13.2 \pm 0.7$ \\
\tableline
\end{tabular*}
\end{table}

\subsection{Extended Tracking and Association with CME Observations}
\label{subsec:fsi_check}

To assess whether the prominence's oscillatory behavior continued or evolved at greater heliocentric distances, we applied the same co-moving slit methodology described in Section~\ref{co-moving slit} to the FSI 304~\AA\ sequence covering the corresponding interval (Figure~\ref{fig:xt_comparison}(b)). Given FSI's substantially larger field of view relative to SUVI, this comparison also tested whether additional oscillation cycles might be recovered at later times or greater heights than were accessible in the SUVI data. Visual inspection of the resulting FSI time-distance map does not show any new oscillation cycles compared to those already identified in SUVI, despite FSI's larger field of view. This might be because of FSI's coarser spatial resolution and pixel scale, which limit the detection of oscillation cycles whose amplitude has likely damped considerably by this stage. We therefore do not draw any conclusion regarding the physical termination of the oscillation from this non-detection. The FSI time-distance map is instead used to confirm the continuity of the eruption into the outer corona, including its connection to the subsequent coronal mass ejection (CME) observed in white-light coronagraph data.

\begin{figure}
\centering
\includegraphics[width=0.95\linewidth]{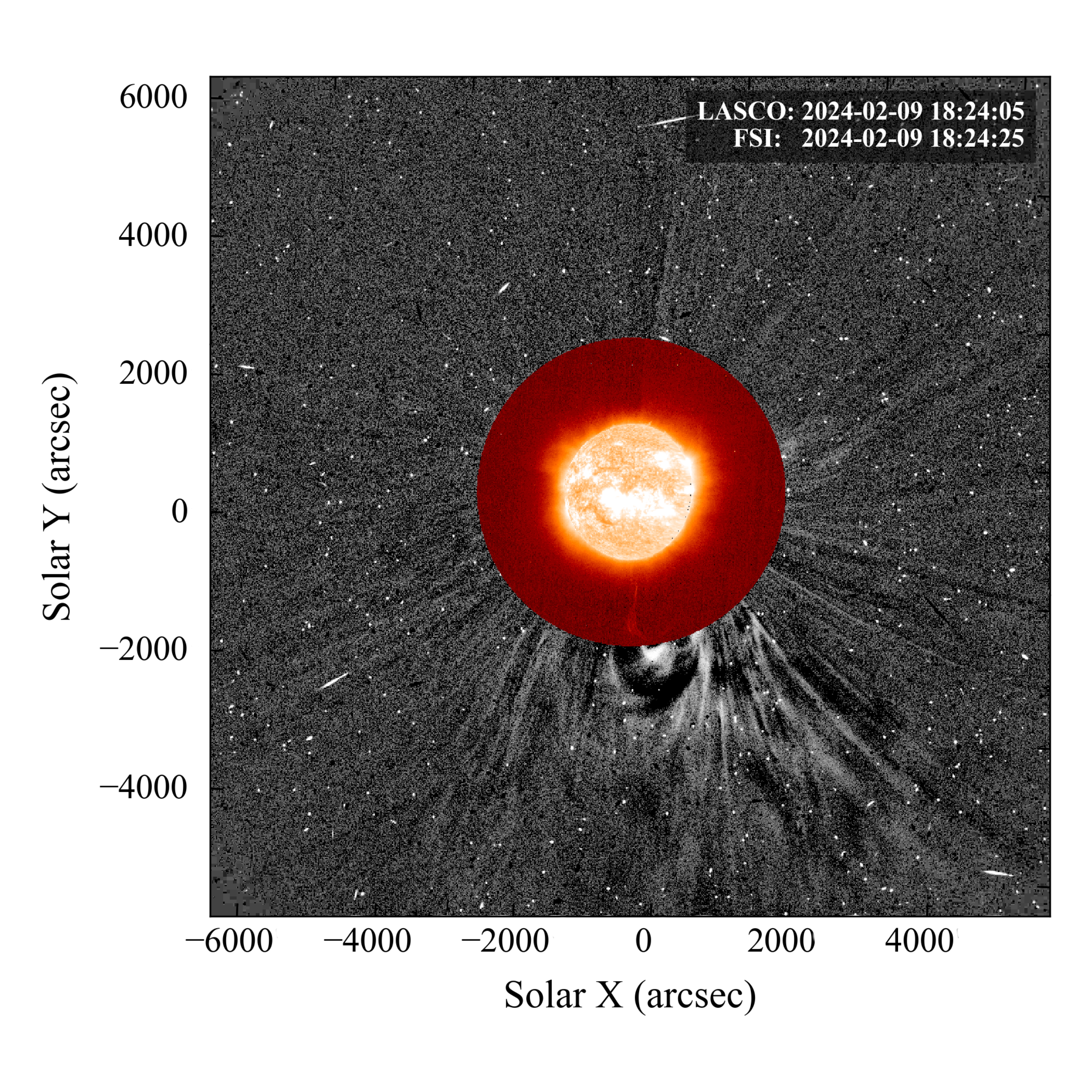}
\caption{A blended view connecting the low-coronal prominence eruption to the subsequent coronal mass ejection (CME). The inner disk shows Solar Orbiter/FSI 304~\AA\ observations of the erupting 
prominence, visible out to $\sim$2.3~R$_{\odot}$ in this frame (within FSI's $\sim$6~R$_{\odot}$ field of view on this date), overlaid on a SOHO/LASCO~C2 running-difference image showing the expanding CME front further out in the corona.}
\label{fig:fig9}
\end{figure}

To visualize this spatial and kinematic connection, we created a composite figure by overlaying extreme-ultraviolet observations of the solar disk from FSI directly onto a near-simultaneous LASCO~C2 running-difference image as shown in Figure~\ref{fig:fig9}. This image shows a clear spatial continuity between the FSI-tracked prominence material and the expanding CME front, confirming that the eruption traced throughout this study is directly associated with this CME.

\begin{figure}
\centering
\includegraphics[width=\columnwidth]{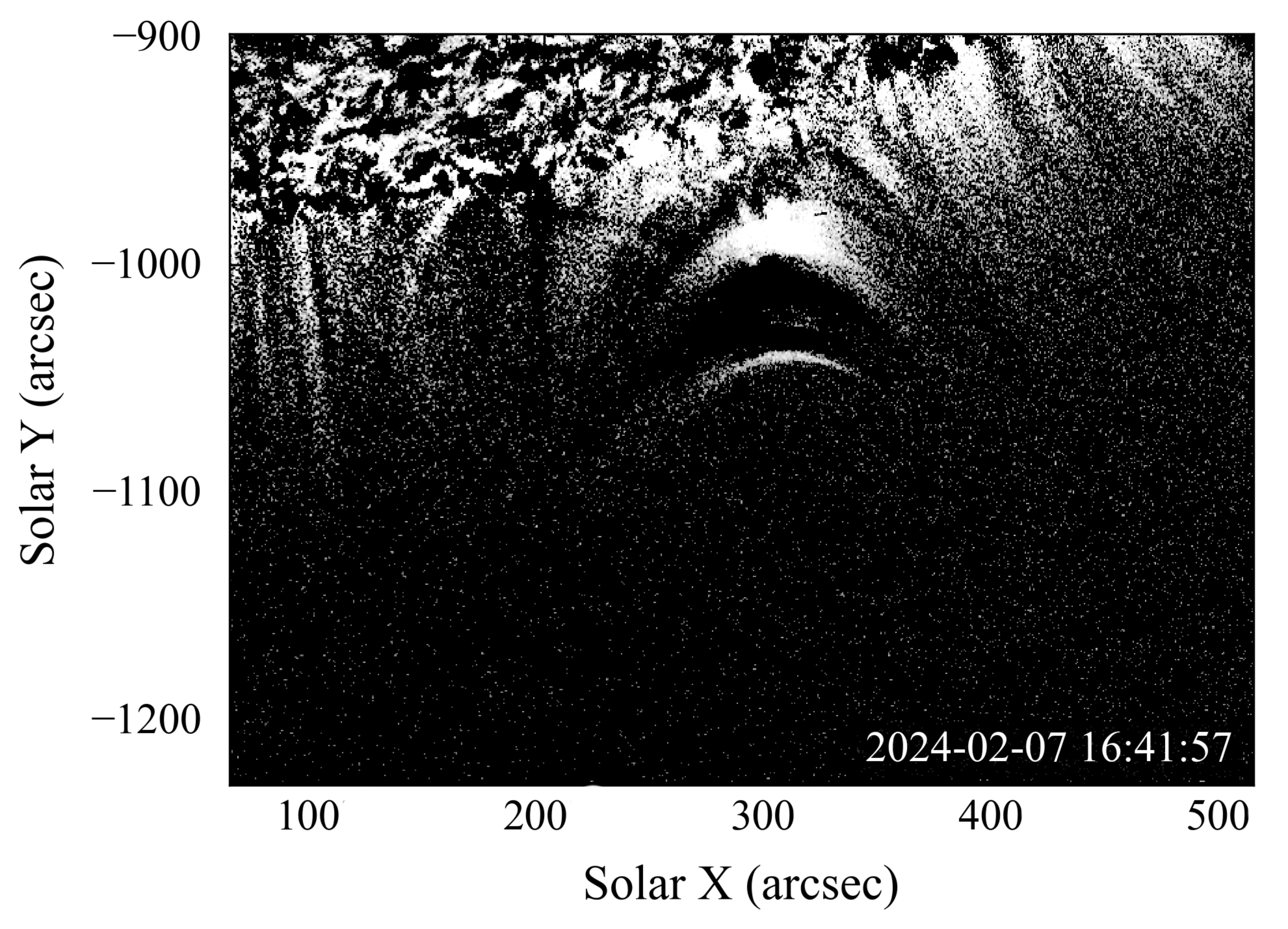}
\caption{AIA 171~\AA\ base-difference image of the same prominence region acquired approximately two days before the oscillation onset, showing the coronal cavity (dark arched structure) within which the prominence is embedded. This suggests that the apparent single-column morphology observed in SUVI 304~\AA\ (Figure~\ref{fig:slit}) reflects a cavity supported structure viewed in projection, rather than a genuinely simple vertical prominence.}
\label{fig:fig11}
\end{figure}

\section{Discussion}
\label{sec:discussion}

\subsection{Physical Interpretation of the Oscillation Evolution}
\label{subsec:interpretation}

When the pre and post-eruption properties are computed, a clear distinction emerges in the oscillation parameters. The observed oscillation period increases across the bifurcation, lengthening from $\sim$34 to $\sim$54~min (Table~\ref{tab:table1}). Concurrently, the oscillation amplitude and the corresponding velocity amplitude exhibit a continuous decrease. The analytic model developed in Section~\ref{subsec:model} offers a qualitative physical framework to explain these observed trends. Based on Equation~(\ref{eq:omega1}), in the limit where $h+z > d$ and vertical displacement $z$ is small, the horizontal oscillation frequency remains nearly constant or decreases very slowly. Consequently, prior to the eruption, the change in the oscillation period is negligible or very slow, which aligns with our pre-eruption observations. However, following the eruption onset, $z$ increases drastically ($h+z \gg d$), and the horizontal oscillation frequency reduces to the form given in Equation~(\ref{eq:12}). This relationship dictates that the period should lengthen significantly post-eruption, scaling proportionally with height (Period $\propto z$). While our observations clearly show the increase in period post-eruption, we do not confirm here a continuous change in period with a linear dependence. Perhaps this is because the post-eruption interval contains barely two oscillation cycles and is fitted with a single, constant period, so any variation of $P$ with height within it cannot be resolved. Furthermore, as depicted in the modeled temporal evolution of the normalized horizontal displacement (Figure~\ref{fig:theory_nonlinear}, middle panel), the theoretical framework qualitatively captures the continuous decrease in oscillation amplitude that occurs alongside the progressively lengthening cycles post-eruption. We do not attempt to interpret the observed damping times within this framework, as the damping is introduced without a physical basis, applied only as a numerical dissipation term to reproduce a damped sinusoidal profile.

There are a few caveats to keep in mind when applying the model this way. First, it describes the bulk transverse motion of a single point mass, so it has nothing to say about the bifurcation into two segments or the mass drainage seen in the real event, these are separate physical processes that a point-mass current loop cannot capture. Second, the model marks the onset of eruptive rise through a rapid vertical acceleration in $z(t)$, which is a dynamical criterion. This is different from the structural criterion we use in the observations, namely the bifurcation of the upper segment, confirmed independently by the associated CME in LASCO C2. These two criteria describe different aspects of the same transition, and we do not expect their timings to coincide exactly.

\subsection{Consistency with Other Theoretical and Numerical Models}
\label{subsec:other_models}

The trend of increasing period with height is not
unique to the model adopted here, and recurs across other independent theoretical and numerical frameworks. The classical \citet{Kuperus1974} model exhibits the same qualitative behavior, since the restoring force weakens with height, the oscillation period correspondingly grows as $P_{\mathrm{KR}} \propto \sqrt{h}$. In the longitudinal oscillation framework, the period instead scales with the radius of curvature of the dipped magnetic field lines \citep{Luna2012}. \citet{Bi2014} reported variations in oscillation periods during successive cycles of longitudinal prominence oscillations, which can be interpreted in terms of changes in the curvature radius of the supporting dipped field lines. Although these theoretical frameworks invoke distinct physical mechanisms for the primary restoring force, such as magnetic tension, field-aligned gravity, or the complex dynamics of full MHD simulations, they yield a consistent qualitative outcome that the oscillation period is sensitive to both prominence height and the progressive destabilization of its supporting structure. This theoretical convergence supports our interpretation of the observed period lengthening as a seismological signature of weakening magnetic support, while recognizing that the present model provides a qualitative rather than unique quantitative explanation.

\subsection{Transverse versus Longitudinal Modes and Seismological estimations}
\label{subsec:los}

The oscillation we have modeled and reported here is the transverse horizontal one. Several of the models discussed above, particularly the curvature-dependent longitudinal framework, predict that a longitudinal mode should lengthen in a similar way, so it is plausible that this prominence supported transverse and longitudinal oscillations at once, with our slit geometry more sensitive to the transverse signal. We cannot confirm or rule out a longitudinal component with the data available.

Projection effects add a further ambiguity. The prominence appears in SUVI 304~\AA\ images as an apparently single, columnar vertical structure (Figure~\ref{fig:slit}). An AIA 171~\AA\ base-difference image of the same region acquired approximately two days before the event (Figure~\ref{fig:fig11}), however, reveals a coronal cavity enclosing the prominence material. The columnar appearance in SUVI/FSI 304~\AA\ is therefore, at least in part, a line-of-sight projection of a more complex, cavity-supported three-dimensional structure viewed close to the plane of the sky. From the available vantage points we cannot determine whether the measured horizontal displacement is purely transverse or contains a longitudinal contribution projected along the line of sight.

The observations of transverse oscillations are typically utilized through models that relate the observed period to the magnetic field strength and the length of the prominence \citep[e.g.,][]{Hyder1966,Kleczek1969}. However, in our case the complex, cavity-supported morphology (Figures~\ref{fig:slit} and~\ref{fig:fig11}) of the prominence and projection effects make the prominence length measurements, and consequently the field strength estimations, highly uncertain.
Numerical simulations by \citet{liakh_numerical_2020,Liakh2023} have shown that transverse perturbations of a magnetic flux rope can excite both transverse oscillations of the magnetic structure and longitudinal oscillations of the cool prominence plasma. If the observed displacement is interpreted as the projection of such a longitudinal oscillation, the pendulum model of \citet{Luna2012} can be utilised to estimate the radius of curvature $R$ of the dip from the relation
\begin{equation}
    P = 2\pi \sqrt{\frac{R}{g_0}}\,,
\end{equation}
where, $g_0 = 274~\mathrm{m~s^{-2}}$ is the acceleration due to gravity. Further, requiring that the magnetic tension at the bottom of the dip, $B^2/(\mu_0 R)$ should support the plasma weight $\rho g_0$, where $\rho = 1.27\,m_\mathrm{p}\,n_e$, one could obtain a lower limit on the field strength  using
\begin{equation}
    B_{min}~[\mathrm{G}] \ge 26 \left( \frac{n_e}{10^{11}~\mathrm{cm}^{-3}} \right)^{1/2} P~[\mathrm{hours}]\,.
\end{equation}
Adopting the typical number density of the prominence ($n_e \approx 1 \times10^{11} ~\mathrm{cm^{-3}}$ ; \citet{Bi2014}) and the observed oscillation period from the pre-eruption phase ($34.3$~min), we obtain the values of $R$ and $B_{\min}$, as 29~Mm and 15~G, respectively. Similarly, the period from the post-eruption phase ($54.3$~min) yields $R = 74$~Mm and $B_{\min} = 23$~G. These values are of the same order as those found previously for large-amplitude longitudinal oscillations \citep{Luna2014,Zhang2017}.

\section{SUMMARY AND CONCLUSIONS}
\label{sec:conclusions}

Using a co-moving slit applied to SUVI and FSI observations of the 2024 February 9 event, we have tracked a single prominence plasma parcel continuously across its entire evolution, from a stable pre-oscillation state through the slow-rise phase, bifurcation into the subsequent eruptive rise (see Table~\ref{tab:timeline} for a summary of the event phases, their timings, and the corresponding instruments used). This continuous tracking of the same material plasma in the pre and post-eruptive datasets is the central methodological contribution of this work and, to our knowledge, has not been demonstrated previously for this type of transition.

The oscillation parameters change systematically across the bifurcation. The period lengthens, the amplitude decays, and the damping time increases, while the associated CME is confirmed independently via LASCO C2. This period lengthening is consistent with the weakening of the restoring force demonstrated by the simplified model in Section~\ref{subsec:model}.
The model further indicates that the observed sequence of oscillation, slow rise and eruption is consistent with a prominence close to the threshold of vertical instability.

Two limitations of the present analysis point directly to future work. First, the current model adopted here treats the prominence as a point mass (a line current), and is thus only strictly applicable to its bulk motion. It does not capture how curvature and mass may vary along an extended body. Extending this framework to an extended prominence structure would allow a more direct, quantitative comparison with the observed height-time and displacement profiles. Second, the identification of the observed motion as purely transverse is subject to line-of-sight ambiguity, given the cavity-supported three-dimensional geometry inferred from the AIA~171~\AA\ base-difference image, multi-vantage-point observations would help resolve this ambiguity in future studies. Additionally, height-dependent variation in gravity may also be considered in future models.

More broadly, because this study is based on a single event, it remains to be established whether the trends observed in the oscillation parameters are general or specific to this particular case. Applying the co-moving slit methodology developed here to a larger sample of events will be necessary to test the generality of this behavior and to assess its potential as a diagnostic tool for validating prominence models, anticipating prominence eruptions and the onset of associated CMEs.

\begin{acknowledgments}
The authors are grateful to the referee for useful comments. The authors thank Vaibhav Pant and Dipankar Banerjee for their valuable advice and discussions during the course of this work. This research has made use of data from the Solar Ultraviolet Imager (SUVI) onboard the GOES-16 satellite provided by NOAA's National Centers for Environmental Information (NCEI). AIA data are courtesy of NASA/SDO and the AIA science team. SDO is a mission of NASA's Living With a Star Program. SOHO is a project of international cooperation between ESA and NASA. The LASCO data used here are produced by a consortium of the Naval Research Laboratory (USA), Max-Planck-Institut f\"{u}r Aeronomie (Germany), Laboratoire d'Astronomie Spatiale (France), and the University of Birmingham (UK). Solar Orbiter is a space mission of international collaboration between ESA and NASA, operated by ESA. The EUI instrument was built by CSL, IAS, MPS, MSSL/UCL, PMOD/WRC, ROB, LCF/IO with funding from the Belgian Federal Science Policy Office (BELSPO/PRODEX); the Centre National d'Etudes Spatiales (CNES); the UK Space Agency (UKSA); the Bundesministerium f\"{u}r Wirtschaft und Energie (BMWi) through the Deutsches Zentrum f\"{u}r Luft- und Raumfahrt (DLR); and the Swiss Space Office (SSO).

SKP is grateful to ISRO for a RESPOND grant (No. ISRO/RES/2/453/25-26). This work benefited from discussions during a visit to the University of Warwick, funded by the Royal Society International Exchanges grant (No. IES/R3/243262).
DK thanks the UKRI Stephen
Hawking Fellowship EP/Z535473/1 and the Latvian Science Council Grant lzp-2024/1-0023. ML acknowledges that this publication is part of the I+D+i project PID2023-147708NB-I00 funded by MICIU/AEI/10.13039/501100011033/ and by FEDER, EU.
\end{acknowledgments}

\bibliography{sample701}{}
\bibliographystyle{aasjournalv7}



\end{document}